\documentclass[journal=jctcce,manuscript=article,layout=twocolumn]{achemso}

\usepackage{graphicx}
\usepackage{subcaption}
\usepackage{pgfplots}
\pgfplotsset{compat=1.18}
\usepackage{xcolor}
\usepackage{multirow}
\usepackage[hidelinks]{hyperref}
\usepackage{braket}
\usepackage{amsmath}
\usepackage{amssymb}
\usepackage{bbold}
\usepackage{booktabs}
\usepackage[exponent-product = \cdot]{siunitx}

\usepackage{pdfpages}
\numberwithin{equation}{section}
\usepackage[version=3]{mhchem}
\usepackage{bm}
\usepackage{dsfont}
\usepackage{cleveref}
\usepackage{notes2bib}
\usepackage{makecell}

\usepackage{tikz}
\usepackage{tikz-feynman}
\tikzfeynmanset{warn luatex=false}
\usetikzlibrary{decorations.markings,arrows.meta}
\usepackage{appendix}

\tikzfeynmanset{compat=1.1.0}
\tikzset{doublePhoton/.style={decorate, decoration={snake, amplitude=.4mm, segment length=1.5mm, pre length=.5mm, post length=.5mm}, double}}
\tikzset{doubleFermion/.style={double,double distance=0.2ex,with arrow=0.5,arrow size=0.15em}}
\tikzset{myFermion/.style={with arrow=0.5,arrow size=0.12em}}
\tikzset{
  myPhoton/.style={
    decorate,
    decoration={snake, amplitude=.4mm, segment length=1.5mm, pre length=.5mm, post length=.5mm},
    line width=1pt
  }
}
\tikzset{feynVertex/.style={circle, fill=blue!30, draw=blue!50, inner sep=1pt}}

\DeclareUnicodeCharacter{2212}{-}

\author{Arno F\"orster}
\email{a.t.l.foerster@vu.nl}
\affiliation{Department of Chemistry and Pharmaceutical Sciences, Vrije Universiteit, De Boelelaan 1105, 1081 HV Amsterdam, The Netherlands}

\title{Third-Order Algebraic Diagrammatic Construction Renormalized by Particle-Particle Ladders}

\begin{document}

\begin{abstract}
The algebraic diagrammatic construction (ADC) of the electronic self-energy is based on M{\o}ller--Plesset perturbation theory. Its vertices contain bare energy denominators that do not capture renormalization of Hartree--Fock excitations through excited-state correlations. As we show here, replacing the M{\o}ller--Plesset amplitudes by amplitudes obtained from Epstein--Nesbet (EN) partitioning of the electronic Hamiltonian overcomes this deficiency at no extra cost by introducing dressed denominators that resum to infinite order the diagonal elements of ladder diagrams described by the pair propagator in the particle-particle $T$-matrix approximation. Third-order ADC with EN-dressed denominators leads to an almost fourfold increase in accuracy compared to standard ADC(3) on a challenging benchmark set for ionization potentials of closed-shell molecules, reaching an accuracy between that of equation-of-motion coupled cluster with singles and doubles (EOM-CCSD) and with singles, doubles, and triples. For open-shell systems, EN-ADC(3) halves the error of ADC(3) for weakly spin-contaminated molecules, outperforming EOM-CCSD, and still improves on ADC(3) for strongly spin-contaminated ones, for which EOM-CCSD remains more accurate.
\end{abstract}

\section{Introduction}
The algebraic diagrammatic construction (ADC) of the electronic self-energy\cite{Schirmer1983} is a M{\o}ller--Plesset (MP)-based perturbation theory (PT)\cite{Chr1934} that, unlike the standard MP$n$ self-energy series,\cite{Cederbaum1975} or diagrammatic expansions based on Hedin's equations,\cite{Hedin1965, Stefanucci2014, Bruneval2025GW+2SOSEXSemidefinite} enforces the correct Lehmann representation of the self-energy at every perturbation order,\cite{Schirmer1983, vonNiessen1984ComputationalFunction} guaranteeing positive semidefinite (PSD) spectral functions.\cite{Winter1972StudyFunction, Cederbaum1975} Third-order ADC [ADC(3)] offers a good compromise between accuracy and computational cost,\cite{Banerjee2021EfficientPairs} yielding ionization potentials (IPs) with errors of the order of 300 meV on average for closed-shell molecules\cite{Dolgounitcheva2016, Opoku2023, Banerjee2023AlgebraicSpectra} at canonical scalings of $\mathcal{O}\left(N^5\right)$ FLOPS and $\mathcal{O}\left(N^4\right)$ memory with system size. 

MP perturbation theory (MPPT) energy denominators vanish whenever orbital energies become closely degenerate, and the amplitudes diverge. This leads to deficiencies of ADC in multi-reference (MR) situations,\cite{Banerjee2023AlgebraicSpectra} a shortcoming that MR-ADC addresses.\cite{Sokolov2018Multi-referenceImplementation,
Chatterjee2019Second-OrderSystems, Chatterjee2020ExtendedExcitations,
Mazin2021MultireferenceBenchmark,
DeMoura2022SimulatingTheory} 
Also in single-reference situations, the MP perturbation series diverges for ground-state energies\cite{Olsen1996SurprisingTheory} and the electronic self-energy.\cite{Hirata2024a} Even at low orders, the accuracy of MPPT quickly deteriorates for strongly polarizable systems. Deficiencies of MP2 or coupled cluster with singles and doubles and perturbative triples [CCSD(T)\cite{Raghavachari1989}] (MPPT on the CCSD wave function\cite{stanton1997ccsd}) for non-covalent interactions of large molecules are prominent examples.\cite{Nguyen2020, Al-Hamdani2021,schafer2025understanding} 
These failures can be traced back to the MP wave function amplitudes not taking into account the interactions between holes and particles in an excited state. Consequently, the energy costs of excitation are underestimated (particles and holes are too close), and their coupling to the ground state overestimated.\cite{Bartlett2007Coupled-clusterChemistry, Fink2016} Renormalizing the MP2 amplitudes by folding higher-order correlation effects, which account for these missing interactions, directly into them, solves this problem.\cite{Masios2023, schafer2025understanding} 

Therefore, it is tempting to investigate whether the accuracy of the ADC of the self-energy can be improved through amplitude renormalization. This work is motivated by the insight that the $G3W2$ self-energy, the next-to-leading term in the expansion of the self-energy in terms of the screened Coulomb interaction, and its 2SOSEX approximation\cite{Bruneval2024} can be made PSD by completing their amplitudes into a square, so that they may be recast as the folds of Hermitian effective Hamiltonians.\cite{Bruneval2025GW+2SOSEXSemidefinite, Marie2026AnSelf-Energy} The resulting equations contain energy denominators, dressed by random phase approximation (RPA)\cite{Macke1950, Gell-Mann1957} excitation energies, in the (outer) scattering vertices that couple the $2n + 1$ ($n=1,2, \dots$) particle sectors to the primary single-particle space. 

Here, we replace the amplitudes in the ADC(3) of the self-energy by amplitudes with dressed energy denominators obtained from the Epstein--Nesbet (EN) partitioning of the electronic Hamiltonian.\cite{epstein1926stark, nesbet1955configuration, Shavitt2009Many-bodyTheory} EN partitioning renormalizes the energy denominators in the ADC amplitudes with approximate single and double excitations.\cite{Jiang2006, Jiang2007, Szabo2012, Fink2016} The double excitations are the diagonal elements of the ladder diagrams that arise from the solution of the particle-particle (pp) $T$-matrix equations,\cite{Orlando2023, Marie2025AnomalousEquation, Marie2025ParquetApproximation, Marie2025Many-bodyChemistry} and describe the repulsion between two excited holes/particles that is missing in the first-order doubles MP energy denominators.
EN-ADC(3) does not aim to be a consistent ADC based on the EN zeroth-order Hamiltonian. We describe the differences in detail in Section~\ref{sec:2.2} and in Appendix~\ref{app:factorization}.

Two works closely connected to the present work have recently been published. Retaining-the-excitation-degree (RE) PT\cite{Fink2006TwoEnergy} pursues a similar goal as EN partitioning but retains, in addition to the diagonal, the off-diagonal interactions between excited configurations of equal excitation degree.\cite{Fink2016} EN partitioning is the diagonal approximation to RE partitioning. Second-order REPT [REPT(2)] has recently been used to formulate an ADC of the polarization propagator that achieved high accuracy at third order.\cite{Leitner2026RE-ADC:Partitioning} Size-consistent Brillouin--Wigner PT,\cite{Keller2022, Carter-Fenk2023RepartitionedEnergy, Dittmer2025RepartitioningModels} whose denominators contain a state-specific correlation energy shift, has likewise been used to formulate an ADC of the polarization propagator.\cite{Dittmer2026BWs-ADC:Theory} 

What motivates the choice of EN partitioning for this work is that RE first-order doubles amplitudes are obtained from solving the iterative linearized coupled cluster doubles (CCD) equations (lin-CCD)\cite{Cizek1966, taube2009rethinking} at $\mathcal{O} \left(N^6\right)$. This is inconsequential for the polarization propagator as its ADC(3) already scales as $\mathcal{O} \left(N^6\right)$, but REPT(2) would raise the formal scaling of ADC(3) to $\mathcal{O} \left(N^6\right)$. We further note that the RE amplitude equations were reported not to converge for several aromatic systems.\cite{Leitner2026RE-ADC:Partitioning} 
EN partitioning is not affected by these issues.

The price to be paid is that EN perturbation theory (ENPT) is size-inconsistent in a canonical orbital basis, even though it remains size-extensive, not invariant with respect to rotations amongst the occupied and virtual orbital manifolds,\cite{Shavitt2009Many-bodyTheory} and the EN zeroth-order Hamiltonian is not SU(2) symmetric. 

Here we show that EN-dressed amplitudes lead to tremendous improvements over standard ADC(3). In Section~\ref{theor}, we introduce EN-ADC(3) and discuss its physical content, before we benchmark its accuracy in Section~\ref{sec:results} for closed-shell and open-shell systems. In Appendix~\ref{appendix}, we point out explicitly the connections to the pp-$T$-matrix approximation.

\section{\label{theor}Theory}
\subsection{M{\o}ller--Plesset and Epstein--Nesbet Amplitudes}
In MPPT, the exact electronic Hamiltonian $\hat{H}$ is partitioned as
\begin{equation}
    \hat{H} = \hat{H}^{\text{MP}}_0 + \hat{V}^{\text{MP}}
\end{equation}
with
\begin{equation}
\label{eq:mpH0}
    \hat{H}_0^{\text{MP}} = \sum_p \epsilon_p \hat{a}^\dagger_p \hat{a}_p
\end{equation}
and 
\begin{equation}
    \hat{V}^{\text{MP}} = \hat{H} - \hat{H}_0^{\text{MP}} \;.
\end{equation}
We consider a Hartree--Fock (HF) ground state throughout, and $\epsilon_p$ denotes a HF orbital energy.
Epstein and Nesbet instead used the diagonal of the exact Hamiltonian in a basis of configurations labeled by $\mu$ as the zeroth-order Hamiltonian,
\begin{equation}
\label{eq:enH0}
    \hat{H}_0^{\text{EN}} = \sum_{\mu} \ket{\Phi_\mu} \langle \Phi_\mu | \hat{H} | \Phi_\mu \rangle \bra{\Phi_\mu} \;,
\end{equation}
so that
\begin{equation}
    \hat{V}^{\text{EN}} = \hat{H} - \hat{H}_0^{\text{EN}} = \sum_{\mu \neq \nu} \ket{\Phi_\mu} \langle \Phi_\mu | \hat{H} | \Phi_\nu \rangle \bra{\Phi_\nu} \;.
\end{equation}
Both partitionings lead to an expansion of the exact ground state $\ket{\Psi} \approx \ket{\Phi^{(0)}} + \ket{\Phi^{(1)}} + \ket{\Phi^{(2)}} + \dots$, where higher-order contributions are linear combinations of excited states of the zeroth-order Hamiltonian, with their weights given by excitation amplitudes $t_\mu$. In this work, only the first-order doubles amplitudes
\begin{equation}
    t_{\mu}^{(1)} = \frac{\langle \Phi_\mu | \hat{V} | \Phi_0 \rangle}{E_0^{(0)} - E_\mu^{(0)}}
\end{equation}
and the second-order singles amplitudes
\begin{equation}
    t_{\mu}^{(2)} = \frac{1}{E_0^{(0)} - E_\mu^{(0)}} \sum_{\nu} \langle \Phi_\mu | \hat{V} | \Phi_\nu \rangle t_{\nu}^{(1)}
\end{equation}
are relevant. In MPPT, 
\begin{equation}
\label{mp_amplitude2}
    t_{ij}^{ab(1)} = \frac{\langle ab || ij \rangle}{\epsilon_i + \epsilon_j - \epsilon_a - \epsilon_b}
\end{equation}
and
\begin{equation}
\label{mp_amplitude1}
\begin{aligned}
    t_{i}^{a(2)} &= \frac{1}{\epsilon_i - \epsilon_a} \Bigg( \frac{1}{2} \sum_{jbc} \braket{aj||bc} t_{ij}^{bc(1)} \\
    &\qquad \qquad\qquad- \frac{1}{2} \sum_{jkb} \braket{jk||ib} t_{jk}^{ab(1)} \Bigg) \;.
\end{aligned}
\end{equation}
$\langle pq||rs \rangle$ denotes an antisymmetrized 4-center integral in Dirac notation.
EN-partitioning instead leads to
\begin{equation}
\begin{aligned}
    E_0^{(0)} - E_{ij}^{ab(0)} &= \langle \Phi_0 | \hat{H} | \Phi_0 \rangle - \langle \Phi_{ij}^{ab} | \hat{H} | \Phi_{ij}^{ab} \rangle \\
    & = \epsilon_i + \epsilon_j - \epsilon_a - \epsilon_b - \Delta_{ij}^{ab} \;,
\end{aligned}
\end{equation}
with (see \citet{Jiang2006})
\begin{equation}
\label{en_ladders}
\begin{aligned}
 \Delta_{ij}^{ab} = & \langle ij|| ij \rangle +
 \langle ab|| ab \rangle 
 - \langle ia|| ia \rangle 
  - \langle jb|| jb \rangle \\
  & 
    - \langle ib|| ib \rangle
    -\langle ja || ja \rangle \;,
    \end{aligned}
\end{equation}
and the EN-amplitudes become
\begin{equation}
\label{en_amplitude2_old}
    \tilde{\tilde{t}}_{ij}^{ab(1)} = \frac{\langle ab || ij \rangle}{\epsilon_i + \epsilon_j - \epsilon_a - \epsilon_b - \Delta_{ij}^{ab}}
\end{equation}
and
\begin{equation}
\label{en_amplitude1_old}
\begin{aligned}
    \tilde{\tilde{t}}_{i}^{a(2)} &= \frac{1}{\epsilon_i - \epsilon_a + \langle ia || ia \rangle } \Bigg( \frac{1}{2} \sum_{jbc} \braket{aj||bc} \tilde{\tilde{t}}_{ij}^{bc(1)} \\
    & \qquad\qquad \qquad \qquad- \frac{1}{2} \sum_{jkb} \braket{jk||ib} \tilde{\tilde{t}}_{jk}^{ab(1)} \Bigg) \;.
\end{aligned}
\end{equation}
The denominators of the EN-singles amplitudes Eq.~\eqref{en_amplitude1_old} are the diagonal elements of the configuration-interaction (CI) singles (CIS) Hamiltonian, while the denominators in Eq.~\eqref{en_amplitude2_old} are the diagonal elements of the CI doubles (CID) Hamiltonian. This makes apparent the interpretation of the EN-amplitudes as being renormalized through excited-state electron-hole interactions. At second order, EN-partitioning introduces an infinite series of pp and particle-hole (ph) ladders into the amplitudes.\cite{Jiang2006, Szabo2012} 

Let us now discuss the physical content of these resummations. Since the energies of the virtual orbitals are (much) higher than those of the occupied ones in a stable HF ground state, the MP-denominators are negative. Physically, this means that the system's HF ground state is energetically well-separated from excited determinants. The terms $\langle ab||ab \rangle$ and $\langle ij||ij \rangle$ represent the repulsion between two particles or two holes and are therefore positive. They are the forward (pp) and backward (hh) time orderings of the same pair propagator in the particle-particle channel (Appendix~\ref{appendix}). Consequently, their presence in Eq.~\eqref{en_amplitude2_old} makes the denominators more negative and mitigates divergences as orbital energy differences approach zero. This is the same physics level-shifting addresses in MR perturbation theories (MRPTs),\cite{roos1995multiconfigurational} and $\kappa$-type regularizers in second-order MP perturbation theories\cite{stuck2013regularized, Lee2018} or quasiparticle self-consistent theories of the electronic self-energy.\cite{Marie2023a, Krieger2026RenormalizationTheory} The four ph terms instead push the denominator closer to zero, exacerbating the problem of diverging energy denominators that plagues MPPT.\cite{Olsen1996SurprisingTheory, leininger2000mo, Hirata2024a} They might even make the denominator positive, which would physically correspond to an excited Slater determinant below the HF ground state.
As we show in Tables~S2 and~S3 of the Supporting Information, the
inclusion of ph ladders turns the doubles denominators of 
BN, C$_2$, BF, BeO, and other molecules positive, and the IPs become physically unreasonable.  
We therefore do not use the amplitudes~\cref{en_amplitude1_old,en_amplitude2_old}
in our EN-ADC calculations to be defined below,
and instead define the modified amplitudes
\begin{equation}
\label{en_amplitude2}
    \tilde{t}_{ij}^{ab(1)} = \frac{\langle ab || ij \rangle}{\epsilon_i + \epsilon_j - \epsilon_a - \epsilon_b - \langle ij|| ij \rangle -
 \langle ab|| ab \rangle} \;,
\end{equation}
that only contain the pp-channel ladder terms (pp and hh), and the correspondingly modified second-order singles amplitudes
\begin{equation}
\label{en_amplitude1}
\begin{aligned}
    \tilde{t}_{i}^{a(2)} &= \frac{1}{\epsilon_i - \epsilon_a + \langle ia || ia \rangle } \Bigg( \frac{1}{2} \sum_{jbc} \braket{aj||bc} \tilde{t}_{ij}^{bc(1)} \\
    & \qquad\qquad \qquad \qquad- \frac{1}{2} \sum_{jkb} \braket{jk||ib} \tilde{t}_{jk}^{ab(1)} \Bigg) \;.
\end{aligned}
\end{equation}
In Appendix~\ref{appendix}, we show that these ladder resummation terms are equivalent to the diagonal of the time-ordered propagator in the pp-$T$-matrix approximation\cite{anderson1958random, Baym1961} (see Refs.~\citenum{Loos2022, Orlando2023, Marie2025AnomalousEquation, Marie2025ParquetApproximation} for modern overviews), also known as Bethe--Goldstone ansatz.\cite{bethe1957effect} This propagator is alternatively obtained by solving the pp-RPA equations\cite{van2013exchange, yang2013double, Yang2015, Zhang2017} that are equivalent to the ladder-CCD equations.\cite{Scuseria2013, peng2013equivalence} Their full solution scales as $\mathcal{O}\left(N^6\right)$, and therefore, the restriction to the diagonal is necessary in a standard density fitting (DF) based implementation if the canonical $\mathcal{O}\left(N^5\right)$ scaling of ADC(3) is not to be raised. We however note that tensor hyper-contraction (THC)\cite{Hohenstein2012, Parrish2012, Hohenstein2012a, Yeh2023} or interpolative separable DF (ISDF)\cite{Duchemin2019, Duchemin2021a, Yeh2024a} could be used to solve the pp-RPA equations in $\mathcal{O}\left(N^4\right)$.\cite{Shenvi2014Tensorr4}

\begin{table}[hbt!]
    \centering
    \begin{tabular}{lll}
    \toprule
     & ring+ladder & ladder only \\
    \midrule
    nonlinear & CCD & \makecell[lc]{ladder-CCD \\ = pp-RPA} \\[6pt]
    linear    & \makecell[lc]{lin-CCD \\ = REPT(2) \\ = CEPA-0} & lin-ladder-CCD \\[6pt]
    diagonal  & ENPT(2) & pp-ENPT(2) \\
    \bottomrule
    \end{tabular}
    \caption{Comparison of several approximate variants of the CCD equations relevant to this work.}
    \label{tab:approaches}
\end{table}

Ladder-CCD is not to be confused with lin-CCD, which in turn is equivalent to REPT(2)\cite{Fink2006TwoEnergy} and also to the coupled electron pair approximation (CEPA)-0 approximation.\cite{meyer1973pno,ahlrichs1979many} We also note that \citet{Carter-Fenk2025DiagrammaticTheory} has recently proposed to restrict linearized CCD to ladders only (lin-ladder-CCD). The relation between these various approaches is summarized in Table~\ref{tab:approaches}. 

\subsection{\label{sec:2.2}Third-order Algebraic Diagrammatic Construction}

Turning now to the ADC of the self-energy, the choice of partitioning of the Hamiltonian only becomes relevant at third order. The ADC(2) self-energy (that is equivalent to the PT2 self-energy) does not contain any amplitudes.\cite{Szabo2012} The ADC(3) self-energy\cite{Schirmer1983} is expressed through the effective Hamiltonian
\begin{equation}
        \label{2,1:effectiveHamiltonian}
\begin{aligned}
    & H^{\text{ADC}(3)}_{\text{eff}}\\
    & = 
\begin{pmatrix}
        \epsilon + \Sigma(\infty) &  U^{I} &  U^{II}  \\ \left(U^{I}\right)^{\dagger} &  {K^{I}} + {C^I} &  0 \\   \left(U^{II}\right)^{\dagger}
         &  0  &  {K^{II}} + {C^{II}} \\ \\
\end{pmatrix} \;.
\end{aligned}
\end{equation}
The first-order doubles and second-order singles amplitudes enter this expression through the static self-energy, given by
\begin{equation}
\label{eq:sigma_static}
    \Sigma_{pq}(\infty) = \sum_{rs} \langle pr || qs \rangle \Delta\gamma^{(2)}_{rs} \;,
\end{equation}
where the second-order residual density is 
\begin{equation}
\label{density}
\begin{aligned}
\Delta\gamma^{(2)}_{pq} = & \frac{1}{2}\Bigg[\sum_{ijabc} \delta_{pb}\delta_{qc}t_{ij}^{ab(1)} t_{ij}^{ac(1)} \\
&
- \sum_{ijkab} \delta_{pi}\delta_{qj} t_{ik}^{ab(1)} t_{jk}^{ab(1)}
\Bigg]   \\
+ & \sum_{bk}\left[\delta_{pb}\delta_{qk} + \delta_{pk}\delta_{qb}\right]t_{k}^{b(2)} \;.
\end{aligned}
\end{equation}
Additionally, the first-order doubles amplitudes enter the outer scattering vertices given by
\begin{equation}
\begin{aligned}
\label{eq:u2_gen_I}
    U^{I,(2)}_{p;qrs} = & \langle pq || rs \rangle-\frac{1}{2} \sum_{ab} t_{rs}^{ab} \langle pq||ab \rangle \\
    & + \hat{P}^-_{sr}\sum_{ja} t_{rj}^{aq} \langle pj||as \rangle \\
    U^{II,(2)}_{p;qrs} = & \langle pq || rs \rangle +\frac{1}{2} \sum_{jl} t_{jl}^{rs} \langle pq||jl \rangle \\
    & - \hat{P}^-_{sr}\sum_{ja} t_{jq}^{ra} \langle pa||js \rangle.
\end{aligned}
\end{equation}
Here, we have defined the operator $\hat{P}_{pq}^- = 1 - \hat{P}_{pq}$, with the permutation operator $\hat{P}_{pq}$ that swaps indices $p$ and $q$. The matrix elements of the three-body blocks are independent of the amplitudes. They contain diagonal terms
\begin{equation}
\begin{aligned}
    K^I_{ija, klb} = & (\epsilon_a - \epsilon_i - \epsilon_j) \delta_{ik} \delta_{jl} \delta_{ab} \\
    K^{II}_{abi, cdj} = & (\epsilon_a + \epsilon_b - \epsilon_i) \delta_{ac} \delta_{bd} \delta_{ij}
\end{aligned}
\end{equation}
that describe the propagation of 3 non-interacting particles, and their first-order interaction in the Faddeev decomposition of the three-particle vertex\cite{faddeev2016scattering, Winter1972StudyFunction, Degroote2011}
\begin{equation}
\label{eq:Ccoupling}
\begin{aligned}
    C^I_{ija, klb} &= \langle ij || kl \rangle \delta_{ab} -  \hat{P}^-_{ij}\hat{P}^-_{kl}\langle ib || ka \rangle \delta_{jl} \\
    C^{II}_{abi, cdj} &= \langle ab || cd \rangle \delta_{ij} - 
     \hat{P}^-_{ab}\hat{P}^-_{cd}\langle aj || ci \rangle \delta_{bd} \;.
\end{aligned}
\end{equation}
The standard ADC(3) Hamiltonian is obtained when the MP-amplitudes~\cref{mp_amplitude2,mp_amplitude1} are used in \cref{density,eq:u2_gen_I}, while EN-ADC(3) uses the pp ladder-restricted EN-amplitudes~\cref{en_amplitude2,en_amplitude1}.
We clarify that EN-ADC(3) is not the ADC that results from applying the ADC procedure to a
perturbation expansion of the self-energy in the EN partitioning. 
Since the EN zeroth-order Hamiltonian contains the diagonal two-body interactions,
its zeroth-order energies are not simple sums of the energies of individual holes and particles.
This is precisely the deficiency of MP partitioning that EN partitioning tries to overcome. 
Consequently, the factorization of time orderings\cite{Shavitt2009Many-bodyTheory} (Appendix~\ref{app:factorization})
that removes the poles at 3h2p and 3p2h energies from the MP self-energy at low orders fails,
and a consistent EN expansion of the self-energy does not take the ADC(3) form that only 
contains 2h1p and 2p1h intermediate states.
Our EN-ADC(3) is therefore understood to be the MP-based ADC(3) in which the amplitudes
have been replaced by EN-dressed ones.

The EN shifts dress the 2p2h ground-state configurations, which enter ADC(3) only through the amplitudes in $U^{I/II,(2)}$ and $\Sigma(\infty)$.
Expanding the EN denominator generates diagonal hh and pp ladder insertions into the ground-state doubles (Appendix~\ref{appendix} and Figure~\ref{fig:ladder_diagrams}).
These diagrams first appear at fourth order in the self-energy. They are not contained in ADC(3), whose higher-order content stems exclusively from resummations within the 2h1p and 2p1h blocks $K+C$.
Therefore, no correlation effects are double counted.
We will also consider a variant of EN-ADC(3) in which we do not include the CIS shift in Eq.~\eqref{en_amplitude1}. The CIS shift is actually attractive, and we will show that this can lead to problems in certain MR situations.

In principle, any amplitudes produced through any of the approaches summarized in Table~\ref{tab:approaches} could replace the first-order doubles MP amplitudes in ADC(3) without any double counting, by the same argument just given: they live in the 2p2h space that is not reached through any of the resummations contained in ADC(3).
Coveney\cite{Coveney2025UncoveringTheory} has recently formulated an IP equation-of-motion (IP-EOM) CCD-type approach in a self-energy framework that bears similarity to ADC(3). Simply plugging the CCD amplitudes into the ADC(3) Hamiltonian is different from his construction. Coveney's construction also modifies the three-particle interaction block Eq.~\eqref{eq:Ccoupling}, which is not affected by the simple amplitude substitutions we propose here, and remains non-Hermitian. For the purpose of rationalizing the behavior of our EN-ADC(3) results, in Section~\ref{sec:results1} we will also show ADC(3) results with CCD amplitudes, which we term CCD-ADC(3). This is in analogy to work by Dreuw and co-workers who used CCD amplitudes in the ADC(2) of the polarization propagator.\cite{Hodecker2019Algebraic-diagrammaticEnergies, Hodecker2019Algebraic-diagrammaticPolarizabilities}

\subsection{Approximate Spin-Adaptation}
Single-particle excitations are necessarily of doublet symmetry, while three-particle excitations can be doublets or quartets. As the MP zeroth-order Hamiltonian is SU(2) symmetric, the effective Hamiltonians of standard ADC are block-diagonal in the basis of configuration state functions (CSF), and can be solved in the doublet manifold. The zeroth-order EN Hamiltonian breaks SU(2) symmetry, and doublet and quartet manifolds therefore remain coupled in EN-ADC(3). Consequently, simply replacing the MP amplitudes in a spin-adapted ADC solver with EN amplitudes introduces an error, as the necessary couplings to the quartet manifold are neglected. To avoid this issue, we define the same-spin (ss) and opposite-spin (os) denominators,
\begin{equation}
\begin{aligned}
    D_{ij, \text{ss}}^{ab} = & \epsilon_i + \epsilon_j - \epsilon_a - \epsilon_b \\
    & - (J_{ij} - K_{ij}) - (J_{ab} - K_{ab}) \\
    D_{ij, \text{os}}^{ab} = & \epsilon_i + \epsilon_j - \epsilon_a - \epsilon_b - J_{ij} - J_{ab} \;,
\end{aligned}
\end{equation}
with $J_{pq} = \langle pq|pq \rangle$ and $K_{pq} = \langle pq|qp \rangle$, and replace the EN-denominators either with spin-summed denominators of the form
\begin{equation}
\label{eq:sum}
\begin{aligned}
    D_{ij, \text{sum}}^{ab} = & \epsilon_i + \epsilon_j - \epsilon_a - \epsilon_b \\
     & - (2J_{ij} - K_{ij}) - (2J_{ab} - K_{ab}) \;,
\end{aligned}
\end{equation}
or spin-averaged denominators
\begin{equation}
\label{eq:averaged}
\begin{aligned}
    D_{ij, \text{average}}^{ab} = & \epsilon_i + \epsilon_j - \epsilon_a - \epsilon_b \\
     & - \big(J_{ij} - \tfrac{1}{2}(1-\delta_{ij})K_{ij}\big) \\
     & - \big(J_{ab} - \tfrac{1}{2}(1-\delta_{ab})K_{ab}\big) \;.
\end{aligned}
\end{equation}
Either of these expressions can replace the MP denominator in a spin-adapted ADC code. Both are pragmatic choices, and neither is a proper spin-adaptation. It is hence not clear whether spin-summation (Eq.~\eqref{eq:sum}) or averaging (Eq.~\eqref{eq:averaged}) leads to a smaller spin-adaptation error. We will therefore compare the spin-adaptation errors induced by both variants in Section~\ref{sec:spin-adaptation-error}.
The shifts only require the integrals $J_{pq}$ and $K_{pq}$, which are obtained from the DF factors at $\mathcal{O}(N_{\text{aux}}N^2)$ cost, so that EN-ADC(3) has the same $\mathcal{O}(N^5)$ scaling and in practice the same cost as standard ADC(3) in a canonical implementation. 

\section{\label{sec:results}Results}

\subsection{Technical Details}

All results are obtained in the Dyson framework, in which the ionization and electron-attachment sectors are treated in the form of the single eigenproblem Eq.~\eqref{2,1:effectiveHamiltonian}.
This is in contrast to the non-Dyson (nD) framework,\cite{Schirmer1998APropagator}
which is often implemented\cite{Dempwolff2019EfficientMolecules, Banerjee2019Third-orderImplementation, Banerjee2021EfficientPairs, Herbst2020Adcc:Methods}
and decouples the ionization and attachment parts of the effective Hamiltonian, so that the latter enters the former only perturbatively.
Unless stated otherwise, we use the $\Sigma(3+)$-scheme (in the nomenclature of Ref.~\citenum{Trofimov2005MolecularApproach}) for the static part of the self-energy, which uses the relaxed MP2/EN2 density matrix in Eq.~\eqref{eq:sigma_static} as opposed to the unrelaxed one that is known to be of relatively poor quality.\cite{Keizer2026BenchmarkMolecules,Bruneval2026GWEquation} For a recent comparison of Dyson- and nD-ADC, as well as for a benchmark of the different approximations to the static self-energy, we refer to Ref.~\citenum{Keizer2026BenchmarkMolecules}. For open-shell systems, we will also employ unrelaxed MP2/EN2 and CCSD density matrices [$\Sigma(3)$ and $\Sigma(\text{CCSD})$ schemes, respectively], the latter being evaluated using pySCF.\cite{Sun2020, Sun2026TheProject} Closed-shell systems are calculated with restricted HF references and spin-adapted equations, and open-shell systems with unrestricted HF (UHF) references in a spin-orbital formulation. All equations have been implemented in matrix-free form in a stand-alone Python package that uses HF orbitals and integrals from pySCF, as well as pySCF's root-following Davidson solver and its coupled-perturbed Hartree--Fock solver needed to relax density matrices. 

We have also implemented IP-EOM coupled cluster with singles, doubles, and triples (IP-EOM-CCSDT) for open-shell systems using the p$^\dagger$q code generator\cite{rubin2021p, Liebenthal2025AutomatedJCTC} to verify the correctness of some of the CCSD(T) reference values for the open-shell systems in Ref.~\citenum{Stahl2022QuantifyingExcitations}.

All (EN)-ADC(3) calculations have been performed using (aug-)cc-pVXZ basis sets, $\text{X = D,T,Q}$,\cite{Dunning1989, Kendall1992, Woon1993} as well as the corresponding (aug-)cc-pVXZ-RI auxiliary basis sets for DF.\cite{Weigend2002, Hattig2005} The precise basis sets used for particular calculations will be discussed together with the results.

The geometries of the oligothiophene chains used to obtain the results presented in Section~\ref{res:pi} have been optimized using the B3LYP\cite{Stephens1994} functional and the def2-TZVP\cite{Weigend2005} basis set, and are included in the Supporting Information.

\subsection{\label{sec:spin-adaptation-error}Spin-Adaptation Errors}

\begin{figure}[hbt!]
    \centering
    \includegraphics[width=\linewidth]{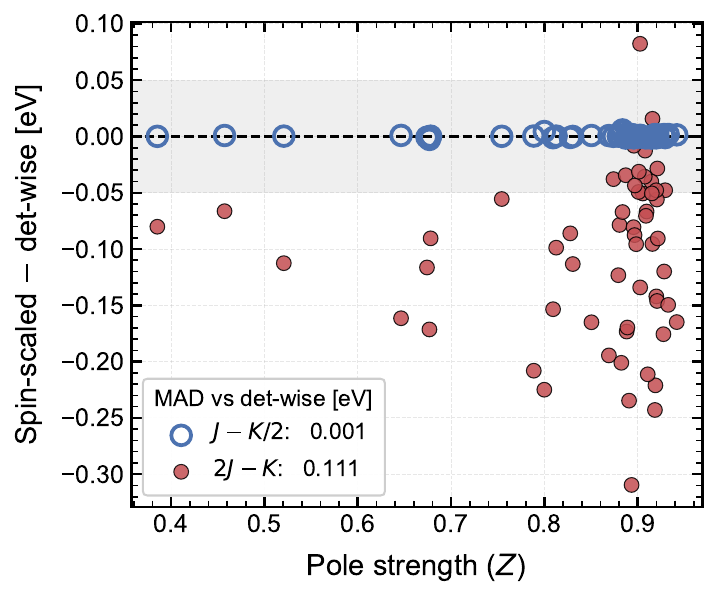}
    \caption{Errors of spin-adapted EN-ADC(3) against the unrestricted (determinant-wise) reference, both evaluated with the coupling of Eq.~\eqref{eq:u2_gen_I}, for spin-averaged (Eq.~\eqref{eq:averaged}, blue markers) and spin-summed (Eq.~\eqref{eq:sum}, red markers) denominators, using an aug-cc-pVTZ basis set.}
\label{fig:en_spin_adaptation_calibration_aug-cc-pvtz}
\end{figure}

We first investigate the spin-adaptation error of the EN amplitudes in EN-ADC(3). To this end, we report in Figure~\ref{fig:en_spin_adaptation_calibration_aug-cc-pvtz} the errors of the two spin-adaptation variants defined in \cref{eq:sum,eq:averaged} against the spin-unrestricted reference. The spin-averaged denominators, Eq.~\eqref{eq:averaged}, reproduce the determinant-wise reference almost exactly, with mean absolute deviations (MADs) of 1~meV and maximum deviations of 6~meV for both the hh and the hh+pp channel. The spin-summed denominators, Eq.~\eqref{eq:sum}, systematically underestimate the IPs, by 0.10~eV (hh) and 0.11~eV (hh+pp) on average, with maximum deviations of 0.63 and 0.31~eV, respectively. 

\subsection{\label{sec:results1}Closed-Shell Systems}
\begin{figure*}[hbt!]
    \centering
    \includegraphics[width=\linewidth]{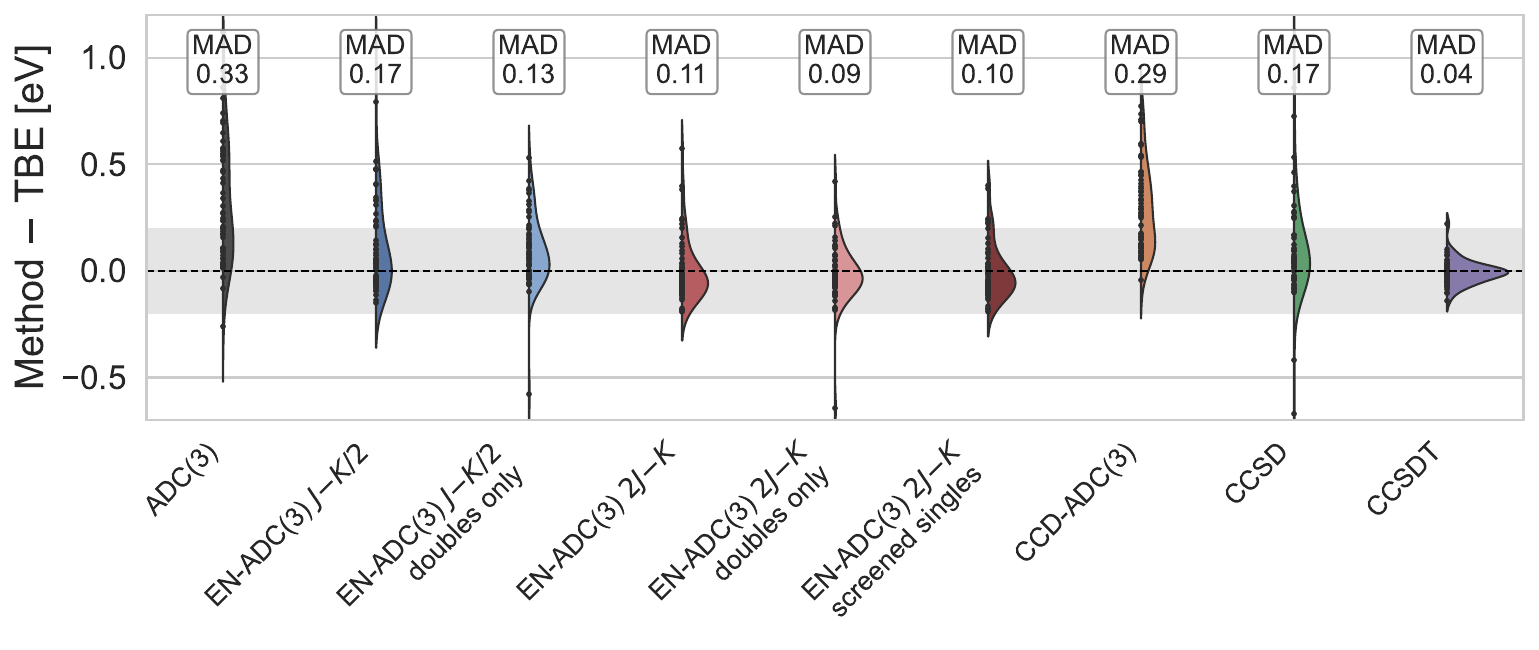}
    \caption{Errors of the IPs (method $-$ TBE) of 23 closed-shell molecules (56 states, aug-cc-pVQZ) for Dyson-ADC(3) (left), different variants of EN-ADC(3), CCD-ADC(3), and the EOM-CCSD and EOM-CCSDT results of Ref.~\citenum{Marie2024}. $J-K/2$ denotes spin-averaged denominators as defined in Eq.~\eqref{eq:averaged}, and $2J-K$ spin-summed denominators as defined in Eq.~\eqref{eq:sum}. ``doubles only'' is without the CIS shift in Eq.~\eqref{en_amplitude1}, and ``screened'' denotes the variant with statically screened CIS shift. For each method, we also show the MAD. All values are in eV.}
    \label{fig:mp2_relaxed_en_comparison_sum}
\end{figure*}

We now benchmark the spin-averaged ($J-K/2$) and spin-summed ($2J-K$) variants of EN-ADC(3) against the theoretical best estimates (TBE) of Ref.~\citenum{Marie2024} for 56 IPs of 23 small molecules, obtained from selected configuration interaction (sCI).\cite{Huron1973,Garniron2019QuantumPrograms} This comparison is performed in the aug-cc-pVQZ basis set. We report results with and without the CIS-shift in the EN-singles amplitudes Eq.~\eqref{en_amplitude1}. We refer to the variant without this shift as ``doubles only''. Our results for these four variants, alongside standard ADC(3) as well as the EOM-CC singles and doubles (EOM-CCSD) and singles, doubles, and triples (EOM-CCSDT) results from Ref.~\citenum{Marie2024} are shown in Figure~\ref{fig:mp2_relaxed_en_comparison_sum}. 

Let us first comment on the performance of the reference methods.
Standard ADC(3) overestimates the IPs, and, with a MAD of 0.326 eV, its performance is rather poor. EOM-CCSD performs significantly better with a MAD of 0.174 eV, but it also shows significant outliers exceeding 1 eV. EOM-CCSDT results are of near-sCI quality, with only a single value exceeding a deviation of 200 meV. 

All variants of EN-ADC(3) improve significantly over standard ADC(3) and outperform, or perform at least as well as, EOM-CCSD. Interestingly, the spin-summed variant ($2J-K$) performs a little better than its spin-averaged counterpart. Since the spin-averaged variant reproduces determinant-wise EN-ADC(3) almost exactly (Section~\ref{sec:spin-adaptation-error}), this advantage stems from the systematic underestimation of the IPs by the spin-summed denominators, which cancels part of the overestimation of standard ADC(3). The ``doubles only'' variant of the spin-summed EN-ADC(3) performs marginally better than its counterpart that includes the singles shift, with a MAD of 0.09 eV. However, it also produces a major outlier with an error of $-0.6$ eV against the sCI reference value for the second IP of BN. 

\begin{table*}[hbt!]
\centering
\caption{IPs (eV) for selected states for which spin-summed ``doubles only'' EN-ADC(3) differs by more than 0.2\,eV from the sCI reference. $Z$ is taken from the EN-ADC(3) spin-summed (doubles-only) eigenvector. sum: spin-summed EN-ADC(3); sum (d2): spin-summed, doubles only; scr: spin-summed with screened singles shift.}
\label{tab:low_z_states}
\begin{tabular}{l c cc c c c c c}
\toprule
& & & & & & \multicolumn{3}{c}{EN-ADC(3)} \\
\cline{7-9}
 IP & $Z$ & sCI & CCSD & CCSDT & ADC(3) & sum & sum (d2) & scr \\
\hline
BN     (2) & 0.51 & 13.707 & 13.720 & 13.808 & 13.891 & 14.089 & 13.063 & 13.935 \\
NH$_3$ (3) & 0.53 & 26.833 & 27.915 & --     & 27.573 & 27.231 & 27.252 & 27.233 \\
CO$_2$ (2) & 0.83 & 17.618 & 18.080 & 17.628 & 18.161 & 17.838 & 17.871 & 17.840 \\
BF     (3) & 0.68 & 20.970 & 21.342 & 20.946 & 21.617 & 21.208 & 21.192 & 21.209 \\
Ne     (2) & 0.91 & 48.340 & 48.494 & 48.270 & 49.200 & 48.585 & 48.552 & 48.584 \\
\bottomrule
\end{tabular}
\end{table*}

This brings us to the fifth variant of EN-ADC(3) in Figure~\ref{fig:mp2_relaxed_en_comparison_sum} on which we have not yet commented. Clearly, the second IP of BN is so sensitive to the inclusion or exclusion of the CIS shift because the electron-hole interaction it describes is exceptionally strong. We therefore investigate a variant of EN-ADC(3) with screened singles, in which we replace the direct electron-hole interaction term in the CIS shift with the statically screened one in the RPA, in the same way as in the $GW$-Bethe--Salpeter equation (BSE) method.\cite{Strinati1988, Rohlfing2000, Onida2002, Blase2018} Indeed, as can also be seen in Table~\ref{tab:low_z_states}, screening reduces the second IP of BN from 14.089 to 13.935 eV, the latter number being in much better agreement with the sCI reference. For all other states, the effect of screening the singles shift is of the order of a few meV only.

Omitting the singles shift altogether drops the IP to 13.063 eV, much below the sCI reference value. We also remind the reader that the singles shift only enters the ADC(3) Hamiltonian through $\Sigma (\infty)$, which speaks to the importance of the proper treatment of this term.

The second IP of BN aside, Table~\ref{tab:low_z_states} also shows the other four states from this set for which spin-summed, doubles-only EN-ADC(3) differs by more than 0.2 eV from the exact reference. These are all semi-valence and semi-core states for which the single-quasiparticle picture tends to break down,\cite{Cederbaum1975, Cederbaum1977} but not all of them have low $Z$-factors. The third state of NH$_3$ is challenging, with CCSD producing an error exceeding 1 eV. Standard ADC(3) performs better. EN-ADC(3) gives further improvements, but it cannot fully close the gap to the sCI reference value. We also comment on the second IP of neon shown in the fifth row of Table~\ref{tab:low_z_states}, which is known to be very challenging for $GW$-based approaches.\cite{Mejuto-Zaera2021, Pavlyukh2026ApproachingSelf-Energies} ADC(3) performs substantially better, and EN-ADC(3) improves agreement with the sCI reference further.

Finally, we mention that EN-ADC(3) performs well for some states for which it produces small $Z$-factors. For instance for the third IP of PH$_3$, the second IP of Ar, and the third IP of CS ($Z$-factors of 0.39, 0.66 and 0.66, respectively), errors remain small, while EOM-CCSD uniformly struggles for these systems, with errors of 0.86, 0.73, and $-0.67$ eV, respectively. 

\begin{figure*}[hbt!]
    \centering
    \includegraphics[width=\linewidth]{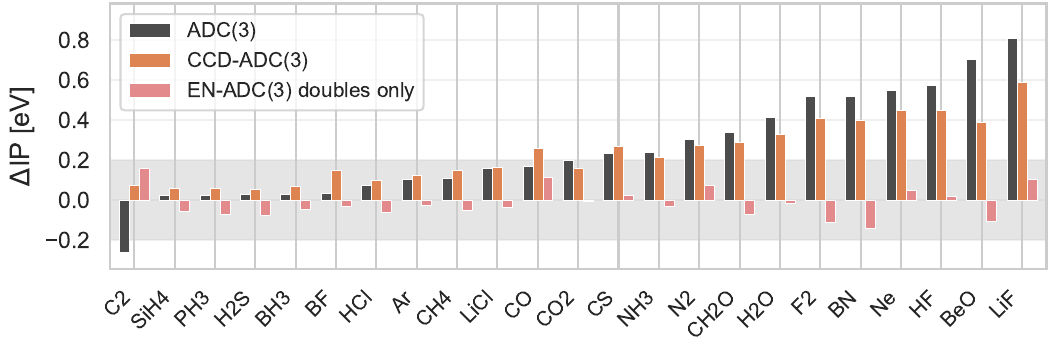}
    \caption{Errors of the first IPs (method $-$ TBE, aug-cc-pVQZ) of ADC(3), EN-ADC(3) (doubles only, $2J-K$), and CCD-ADC(3) for the 23 molecules considered in this work. All values are in eV.}
    \label{fig:cc_adc3_per_system_homo}
\end{figure*}

As established by the discussion around Table~\ref{tab:approaches}, EN amplitudes are diagrammatically contained in the CCD amplitudes. A natural question is therefore how CCD amplitudes affect the performance of ADC(3). The 7th violin in Figure~\ref{fig:mp2_relaxed_en_comparison_sum} shows the performance of CCD-ADC(3). In agreement with the ``doubles only'' variant of EN-ADC(3), the singles amplitudes in the static part of $\Sigma(\infty)$ are left at the MP level. With a MAD of 0.29 eV, this variant performs only marginally better than ADC(3). This indicates that regularization through the EN shift, and not necessarily the quality of the amplitudes, is the mechanism that leads to the stellar performance of EN-ADC(3). This also suggests that the good performance of EN-ADC(3) is due to favorable error cancellation. 

This interpretation is confirmed on a per-system basis with Figure~\ref{fig:cc_adc3_per_system_homo}, which shows the errors of ADC(3), CCD-ADC(3), and EN-ADC(3) for the first IPs of the 23 considered molecules, ordered by increasing error of ADC(3). CCD-ADC(3) tends to marginally increase the errors for the systems for which ADC(3) performs well, and leads to modest improvements for the systems where the ADC(3) errors are large. There is however little correlation with EN-ADC(3) whose performance is uniformly excellent (MAD of 0.06 eV for this subset) across all 23 molecules.

\begin{table}[hbt!]
\centering
\caption{Pearson ($r$) and Spearman ($\rho$) correlation of the ADC(3) errors and of the EN corrections [ADC(3) $-$ EN-ADC(3), doubles only, $2J-K$] of the 23 first IPs (aug-cc-pVQZ) with descriptors evaluated at the HF/aug-cc-pVQZ level: the HOMO--LUMO gap $\Delta\epsilon_{\text{HL}}$, the largest MP2 doubles amplitude $t_{\max}$, and the self-repulsion $J_{\text{HOMO}}$ of the ionized orbital.}
\label{tab:descriptors}
\setlength{\tabcolsep}{4pt}
\begin{tabular}{lcccc}
\toprule
 & \multicolumn{2}{c}{ADC(3) error} & \multicolumn{2}{c}{EN correction} \\
\cmidrule(lr){2-3} \cmidrule(lr){4-5}
 & $r$ & $\rho$ & $r$ & $\rho$ \\
\midrule
$\Delta\epsilon_{\text{HL}}$ & $+0.24$ & $+0.21$ & $+0.15$ & $+0.07$ \\
$t_{\max}$ & $-0.14$ & $+0.17$ & $-0.18$ & $+0.11$ \\
$J_{\text{HOMO}}$ & $+0.71$ & $+0.85$ & $+0.57$ & $+0.72$ \\
\bottomrule
\end{tabular}
\end{table}

\begin{figure}[hbt!]
    \centering
    \includegraphics[width=\linewidth]{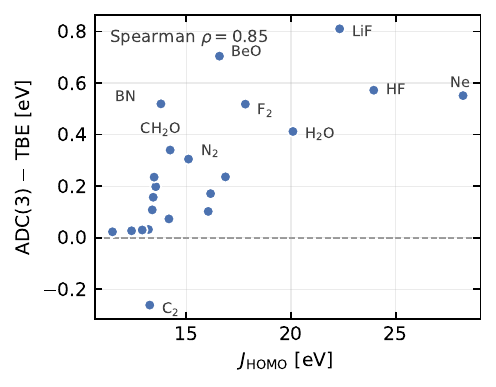}
    \caption{Errors of the first IPs (ADC(3) $-$ TBE, aug-cc-pVQZ) of the 23 closed-shell molecules against the self-repulsion $J_{\text{HOMO}}$ of the ionized orbital, evaluated at the HF/aug-cc-pVQZ level. All values are in eV.}
    \label{fig:adc3_error_vs_jhomo}
\end{figure}

To gain insight into these trends, we show in Table~\ref{tab:descriptors} how the ADC(3) error correlates with a few descriptors that are easy to obtain.
One would expect to observe a correlation with a simple descriptor of MR character,
but the ADC(3) error is uncorrelated with the HF HOMO--LUMO gap (Pearson $r = 0.24$) and with the largest MP2 doubles amplitude ($r = -0.14$).
Instead, it correlates strongly with the self-repulsion $J_{\text{HOMO}}$ of the ionized orbital (Spearman $\rho = 0.85$, Figure~\ref{fig:adc3_error_vs_jhomo}).
Interestingly, this is exactly the quantity the Edmiston--Ruedenberg localization scheme maximizes.\cite{Edmiston1965a}
$J_{\text{HOMO}}$ is directly related to the compactness of the hole left behind after ionization, which also governs the orbital relaxation upon ionization.\cite{Pickup1973DirectShells}
The largest errors occur for compact holes on electronegative atoms (LiF, BeO, HF, Ne, F$_2$, H$_2$O). The EN correction, ADC(3) $-$ EN-ADC(3), follows the ADC(3) error closely ($r = +0.96$),
so that EN-ADC(3) corrects ADC(3) most where it is least accurate.
We also find that restricting the EN shift to the hole--hole terms $\braket{ij||ij}$ (doubles only, $2J-K$) yields about 80\% of the full EN correction ($r = 1.00$).

This observation confirms the intuitive picture outlined in the introduction.
MP amplitudes neglect the repulsion between the two holes of a doubly excited configuration,\cite{Fink2016, Carter-Fenk2023RepartitionedEnergy} which is largest for compact occupied orbitals.
ADC(3) therefore overestimates the IPs most for compact holes, as measured by $J_{\text{HOMO}}$.
The EN hole--hole shift removes this error.

\subsection{Open-Shell Systems}

\begin{figure*}[hbt!]
    \includegraphics[width=\textwidth]{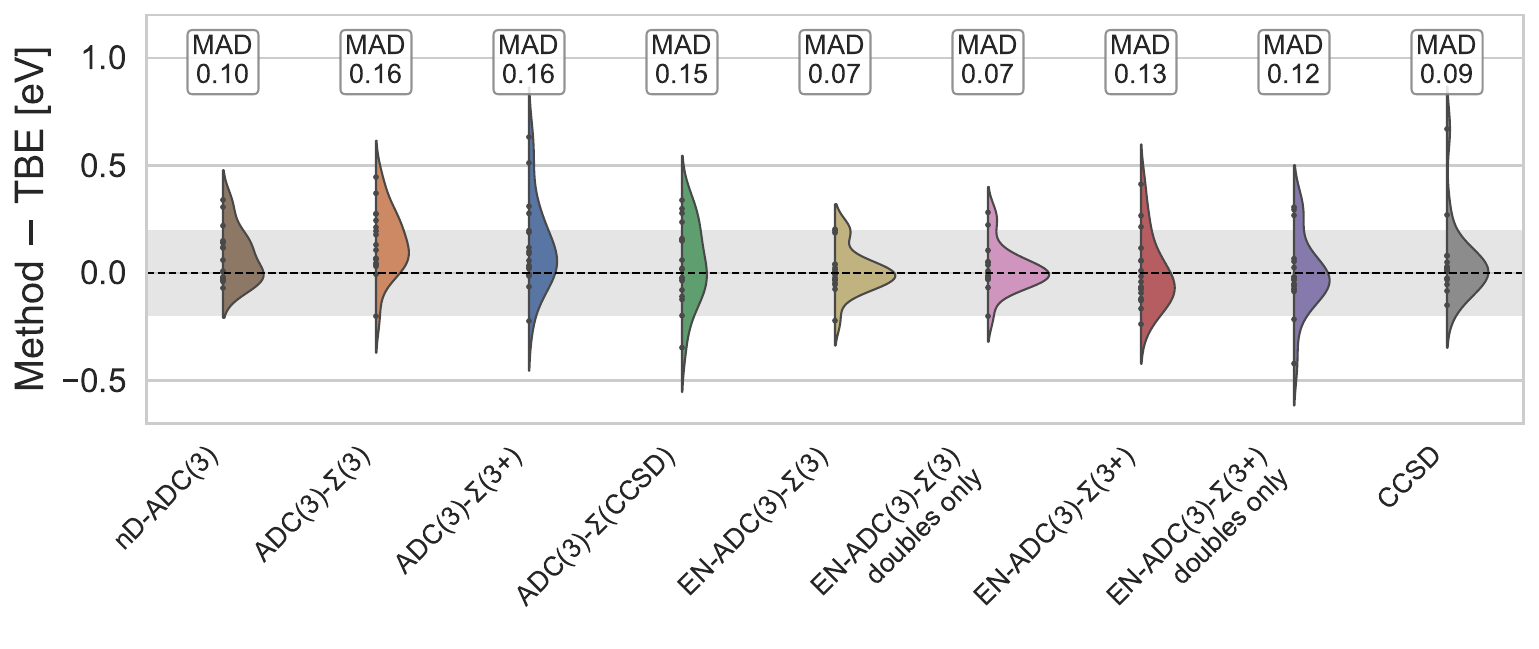}
    \caption{Errors of the first IPs (method $-$ TBE, aug-cc-pVDZ) for standard- and EN-ADC(3) with different approximations to the static self-energy for the set of weakly spin-contaminated open-shell systems of Ref.~\citenum{Stahl2022QuantifyingExcitations}. The TBE is IP-EOM-CCSDT/ROHF where available and CCSD(T)/ROHF otherwise. $\Sigma(3)$, $\Sigma(3+)$, and $\Sigma(\text{CCSD})$ denote the static self-energy evaluated with the unrelaxed MP2/EN2, the relaxed MP2/EN2, and the CCSD density matrix, respectively. EOM-CCSD and nD-ADC(3) results taken from Ref.~\citenum{Stahl2022QuantifyingExcitations} are shown for comparison. All calculations have been performed using an unrestricted HF reference. For each method, we also show the MAD. All values are in eV.}
    \label{openshell_violin_wsm}
\end{figure*}

Next, we investigate the performance of standard and EN-ADC(3) for first IPs of the 40 open-shell systems from Ref.~\citenum{Stahl2022QuantifyingExcitations}, which is divided into 18 weakly spin-contaminated molecules (defined by $\Delta\hat{S}^2 < 0.1$) and 22 strongly spin-contaminated molecules. The reference values have been calculated by \citet{Stahl2022QuantifyingExcitations} using CCSD(T) with a restricted open-shell HF (ROHF) reference and in the aug-cc-pVDZ basis set. Following \citet{Stahl2022QuantifyingExcitations}, we benchmark our UHF-based ADC results against these ROHF-based reference values. For this work, we also calculated IP-EOM-CCSDT/ROHF reference values for the systems for which we could afford it, to assess the quality of the CCSD(T) reference. We indeed found that the CCSD(T) value for the IP of CN of 15.28 eV is probably incorrect. IP-EOM-CCSDT gives 13.90 eV, in very good agreement with EOM-CCSD (13.84 eV) and ADC(3). For most other systems, the deviations of CCSD(T) from IP-EOM-CCSDT are of the order of a few tens of meV at most. Nevertheless, for all systems for which they are available, we replaced the CCSD(T) reference values with IP-EOM-CCSDT. We refer to the Supporting Information for details. 

Figure~\ref{openshell_violin_wsm} shows the results for the set of weakly spin-contaminated systems. As already pointed out by \citet{Stahl2022QuantifyingExcitations}, nD-ADC(3) performs very well here with a MAD of 0.1 eV, on par with EOM-CCSD. Interestingly, operating in the Dyson framework deteriorates this good performance somewhat, leading to a MAD of 0.16 eV for the ADC(3)-$\Sigma(3+)$ method. To rule out that this mismatch is due to the relaxation of the density matrix that is used to evaluate the static self-energy (nD-ADC(3) uses an unrelaxed density matrix), we also checked the $\Sigma(3)$ and $\Sigma(\text{CCSD})$ schemes, but those do not alter the results substantially. 

\begin{figure*}[hbt!]
    \includegraphics[width=\textwidth]{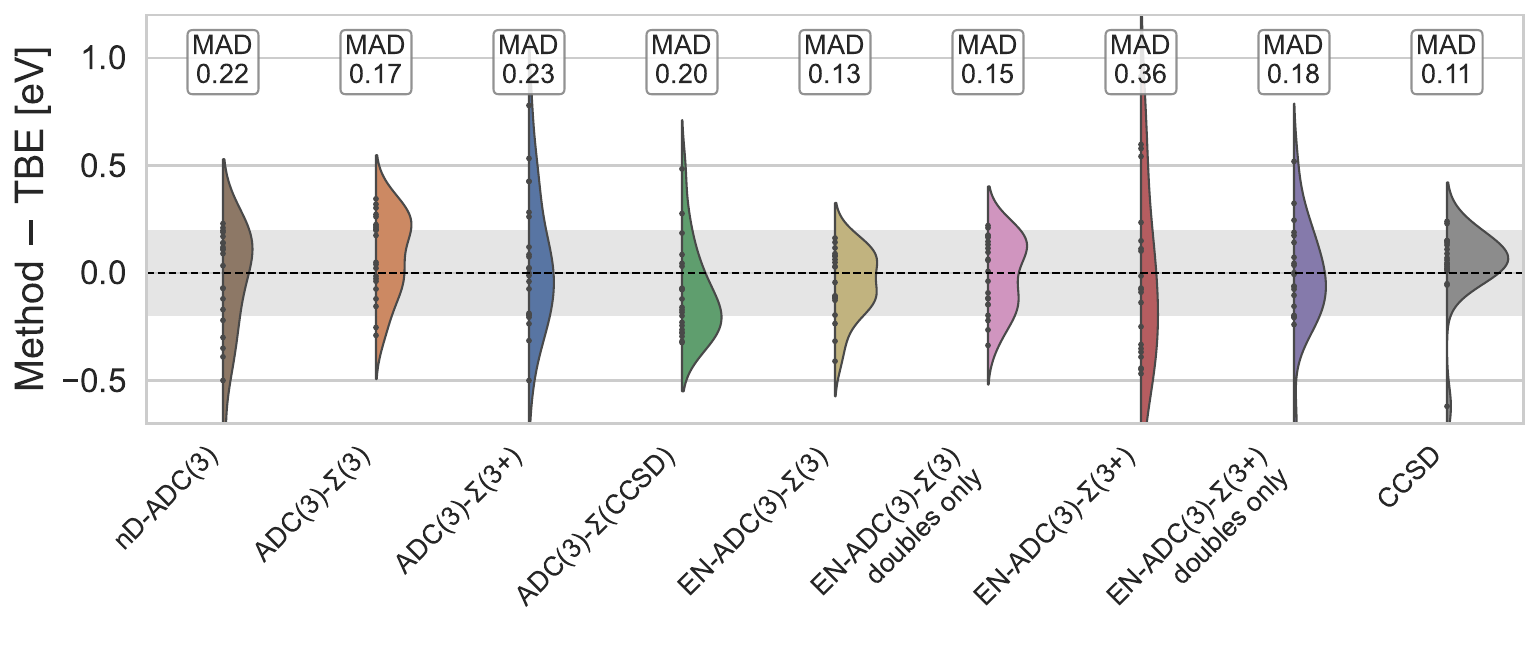}
    \caption{Errors of the first IPs (method $-$ TBE, aug-cc-pVDZ) for standard- and EN-ADC(3) with different approximations to the static self-energy for the set of strongly spin-contaminated open-shell systems of Ref.~\citenum{Stahl2022QuantifyingExcitations}. The TBE is IP-EOM-CCSDT/ROHF where available and CCSD(T)/ROHF otherwise. $\Sigma(3)$, $\Sigma(3+)$, and $\Sigma(\text{CCSD})$ denote the static self-energy evaluated with the unrelaxed MP2/EN2, the relaxed MP2/EN2, and the CCSD density matrix, respectively. EOM-CCSD and nD-ADC(3) results taken from Ref.~\citenum{Stahl2022QuantifyingExcitations} are shown for comparison. All calculations have been performed using an unrestricted HF reference. For each method, we also show the MAD. All values are in eV.}
    \label{openshell_violin_ssm}
\end{figure*}

Turning now to EN-ADC(3)-$\Sigma(3+)$, we find the improvement over standard ADC(3)-$\Sigma(3+)$ to be modest. However, EN-ADC(3)-$\Sigma(3)$ performs very well with a MAD of 0.07 eV, slightly outperforming EOM-CCSD. This result is independent of the inclusion of the CIS-shift in the second-order singles amplitudes Eq.~\eqref{en_amplitude1}. In summary, for weakly spin-contaminated open-shell systems, EN-ADC(3)-$\Sigma(3)$ halves the error of standard ADC(3)-$\Sigma(3)$ (MAD of 0.16 eV) and outperforms EOM-CCSD. 

The smaller errors of standard ADC(3) for the open-shell than for the closed-shell set can be linked to the observation discussed above that 
ADC(3) errors correlate with the self-repulsion of the ionized orbital and thus the compactness of the hole left behind.
The open-shell set consists of radicals that are mostly ionized from the singly occupied orbital and contains none of the compact, ionic, or rare-gas holes that dominate the closed-shell errors.
Consistently, the EN correction is almost three times smaller on average for the open-shell set ($0.10$ eV) than for the first IPs of the closed-shell set ($0.27$ eV). 

We now turn to the more challenging set of strongly spin-contaminated systems in Figure~\ref{openshell_violin_ssm}. Strong spin-contamination often indicates a strongly correlated ground state.  Consequently, nD-ADC(3) performs much worse, with a MAD of 0.22 eV. Dyson-ADC(3) comes with comparable MADs of 0.17, 0.23, and 0.20 eV for the $\Sigma(3)$, $\Sigma(3+)$ and $\Sigma(\text{CCSD})$ methods, respectively. The picture is similar to that for the closed-shell systems discussed previously. These errors are about twice as large as the ones of EOM-CCSD, with a MAD of 0.11 eV. We note that \citet{Stahl2022QuantifyingExcitations} have shown that ROHF reference orbitals lead to substantial improvements over UHF orbitals, but we have not implemented this variant in our ADC code.

EN-ADC(3) achieves improvements over standard ADC(3), but the inclusion of the CIS shift in the second-order singles amplitudes Eq.~\eqref{en_amplitude1} proves problematic in combination with orbital relaxation: with the CIS singles shift, EN-ADC(3)-$\Sigma(3+)$ has a MAD to the TBE reference of 0.36 eV, that drops to 0.18 eV if the singles shift is excluded. We have already observed similar issues for some of the more strongly correlated closed-shell systems discussed in Section~\ref{sec:results1}, confirming that the singles shift is hardly beneficial and might be detrimental to the accuracy of EN-ADC(3). However, these issues come into play only through orbital relaxation. In EN-ADC(3)-$\Sigma(3)$, singles have even a beneficial effect. 

Overall, EN-ADC(3)-$\Sigma(3)$ still improves on standard ADC(3)-$\Sigma(3)$ for strongly spin-contaminated systems (MADs of 0.13 and 0.17 eV, respectively), but for this benchmark set, EOM-CCSD (0.11 eV) remains more accurate.

\subsection{\label{res:pi}$\pi$-Conjugated Molecules}
Before concluding this work, we assess whether the accuracy gains from EN partitioning persist for larger, more extended systems, and whether the magnitude of the denominator shift changes significantly as canonical orbitals delocalize over an increasing number of atoms. We performed all calculations in this subsection using the cc-pVTZ and cc-pVQZ basis sets, with results extrapolated to the complete-basis-set (CBS) limit using the cubic ansatz of Ref.~\citenum{Halkier1998}. We observed that the basis set error was always constant for the first couple of acenes and oligothiophenes. Therefore, for the larger members of each series, we only performed cc-pVTZ calculations and applied the difference to the CBS limit obtained in the calculations for the smaller systems as a correction. In the following figures, all filled data points denote systems for which a cc-pVQZ calculation was performed.

\begin{figure}[hbt!]
    \centering
    \includegraphics[width=\linewidth]{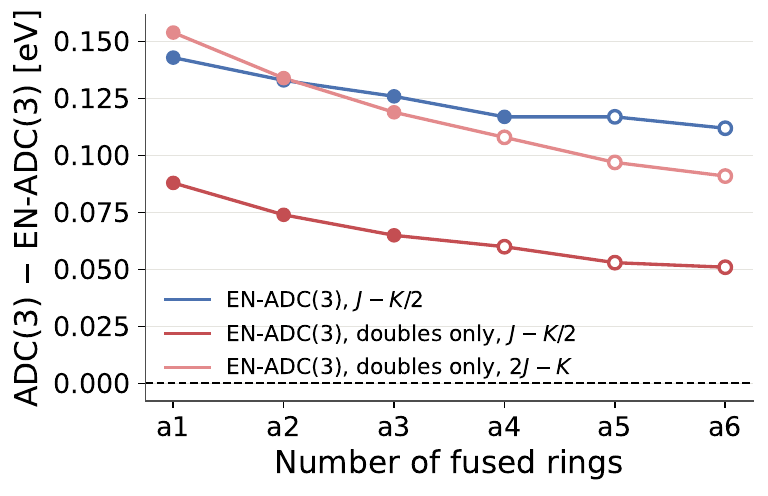}
    \caption{Size of the EN dressing correction, standard ADC(3) minus EN-ADC(3), for the singles-included and both doubles-only variants across the acene series.}
    \label{fig:acenes_dressing}
\end{figure}

\begin{figure}[hbt!]
    \centering
    \includegraphics[width=\linewidth]{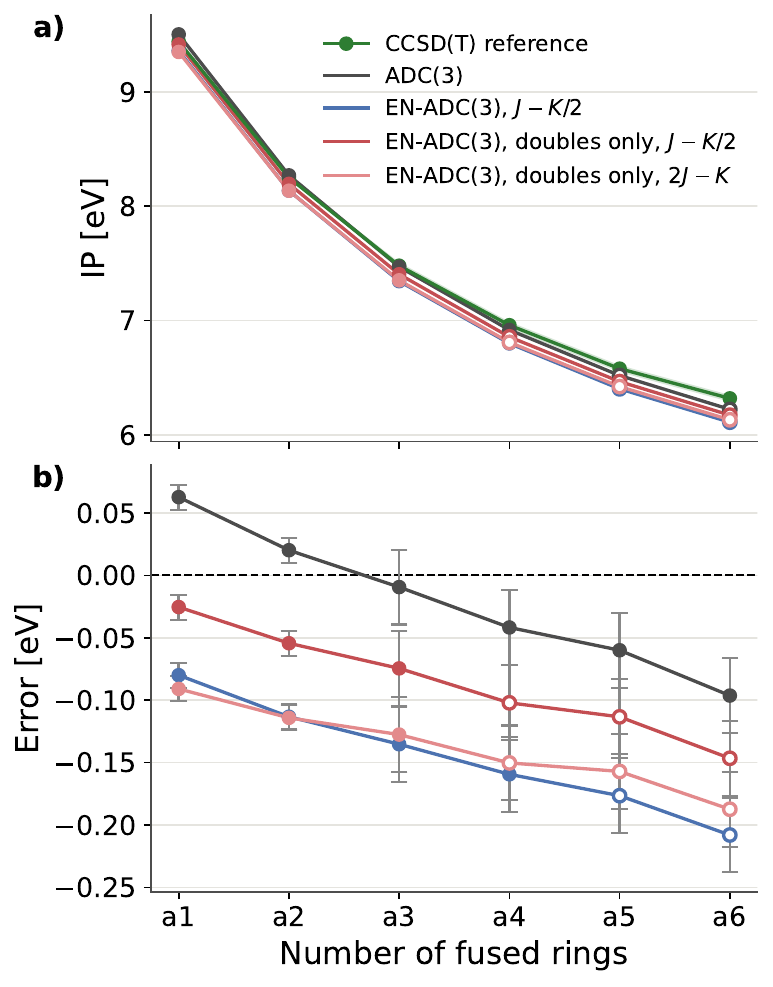}
    \caption{Vertical IPs of the acene series (a1--a6) from ADC(3) and EN-ADC(3) [with singles, and doubles-only, spin-averaged ($J-K/2$) and spin-summed ($2J-K$)], (a) Raw IPs; (b) error relative to CCSD(T).}
    \label{fig:acenes_ip}
\end{figure}

As a first test, we calculate the first IPs of the members of the linear acene series from benzene to hexacene. This series is well-investigated, and many works have calculated the singlet-triplet gap\cite{Hachmann2007, Hajgato2009, Hajgato2011, Zimmerman2017, Ghosh2017, Mostafanejad2019, Meitei2021, Dey2022, Dhingra2023} and electron affinities and IPs\cite{Deleuze2003TheTheories, Deleuze2003BenchmarkOligoacenes,Hajgato2009,Dupuy2015VerticalAnsatz,Rangel2016} as a function of acene size. Here, reference IPs calculated at CCSD(T) and optimized molecular structures are taken from Ref.~\citenum{Rangel2016}. As shown in Figure~\ref{fig:acenes_dressing}, the magnitude of the EN shift decreases moderately with increasing chain length, but remains sizable even for the largest acene considered here (42 atoms). Interestingly, the effect is smaller when the CIS shift is excluded, probably because the electron-hole interaction is rather long-ranged. As shown in Figure~\ref{fig:acenes_ip}, ADC(3) itself changes character along the series, overestimating the reference IPs of the smallest acenes but underestimating them from anthracene (a3) onward. Because the EN shift acts in a fixed direction, it improves agreement with the reference for benzene only but increases the already-negative error further along the series. 

\begin{figure}[hbt!]
    \centering
    \includegraphics[width=\linewidth]{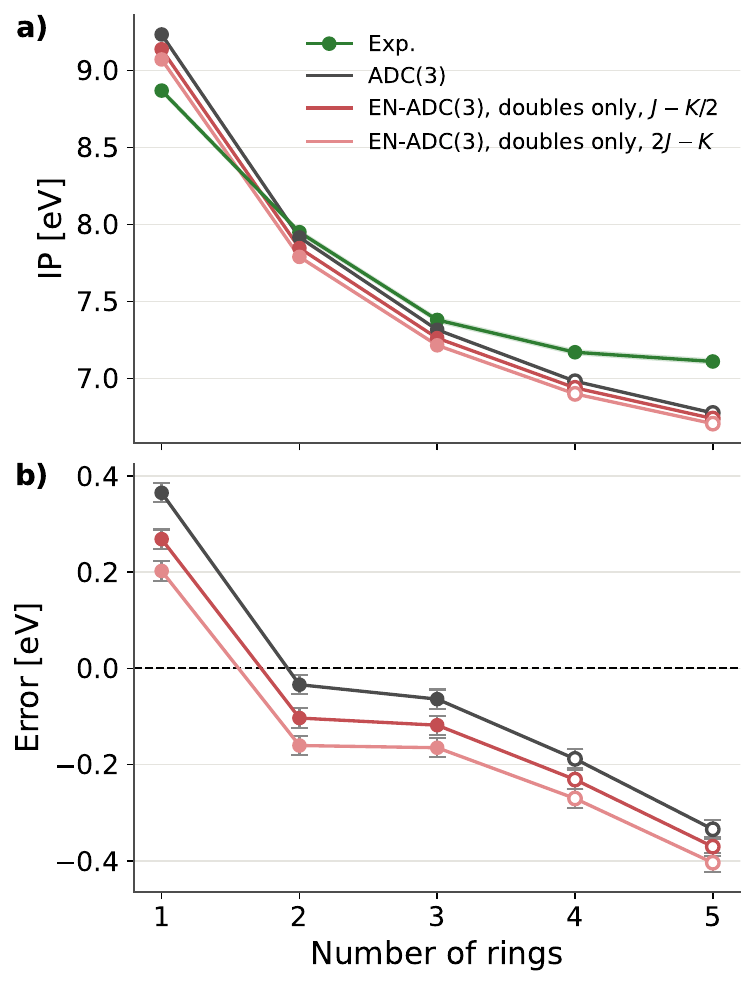}
    \caption{Vertical IPs of the oligothiophene series from ADC(3) and EN-ADC(3) (doubles-only $J-K/2$ and $2J-K$), (a) Raw IPs; (b) error relative to experiment.}
    \label{fig:oligothiophene_ip}
\end{figure}

\begin{figure}[hbt!]
    \centering
    \includegraphics[width=\linewidth]{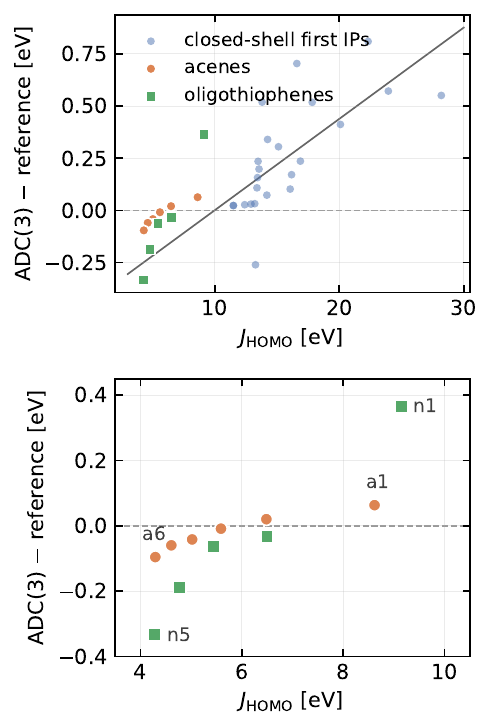}
    \caption{ADC(3) error of the acenes a1--a6 (CBS limit, relative to CCSD(T)) and the oligothiophenes n1--n5 (CBS limit, relative to experiment) against the self-repulsion $J_{\text{HOMO}}$ of the ionized orbital (HF/cc-pVTZ). (a) Together with the 23 closed-shell first IPs of Figure~\ref{fig:adc3_error_vs_jhomo} (relative to the TBE) and their linear fit; (b) enlarged view of the two series.}
    \label{fig:acenes_error_vs_jhomo}
\end{figure}

Next, we investigate the oligothiophene series. As we are not aware of any published theoretical reference values, we compare our results against vertical photoelectron spectroscopy data.\cite{Jones1990Determination, DaSilvaFilho2007Hole-vibronicEnergies} Our results are shown in Figure~\ref{fig:oligothiophene_ip} and display the same qualitative behavior as the acenes: ADC(3) overestimates the IP of thiophene itself but underestimates those of the larger oligomers, and EN-ADC(3) improves upon ADC(3) only for the smallest ring count. We caveat that our calculations do not account for thermal averaging or vibronic couplings, both of which might at least partially explain the large errors of both EN- and standard ADC(3).

Taken together, the acene and oligothiophene series show the same qualitative behavior. For smaller systems, standard ADC(3) overestimates the IPs, but increasingly underestimates them with growing system size. Consequently, the EN shift, which always decreases the IPs of weakly correlated systems, improves agreement with the reference where ADC(3) overestimates, but increasingly worsens it once ADC(3) begins to underestimate.

Whether this crossover behavior holds generally is an interesting open question that broader benchmarks will need to address.
However, as shown in Figure~\ref{fig:acenes_error_vs_jhomo} and consistent with the previous observation for the IPs of the closed-shell molecules discussed in Section~\ref{sec:results1},
the ADC(3) error
correlates with the self-repulsion $J_{\text{HOMO}}$ of the ionized orbital.
$J_{\text{HOMO}}$ (HF/cc-pVTZ) falls from 8.6~eV (benzene) to 4.3~eV (hexacene)
and from 9.1~eV (thiophene) to 4.3~eV (quinquethiophene). 
In particular, these values are well below the range of 11.5--28.2~eV found for the 23 molecules in the benchmark set of \citet{Marie2024}. 
This can be explained by the fact that the electron is removed from the delocalized conjugated $\pi$-system.

\section{Conclusions}

M{\o}ller--Plesset-based perturbation theories estimate the energy cost of an $n$-body excitation as the energy difference between $n$ non-interacting particles and holes. As a consequence, the ADC of the self-energy systematically overestimates the coupling of doubly excited configurations to the single-particle reference state. Renormalizing the amplitudes that enter the ADC(3) self-energy offers a direct and inexpensive route to correcting this deficiency. EN partitioning provides perhaps the simplest realization of this idea, requiring no integrals beyond those already computed in a standard ADC(3) calculation.

We have benchmarked the resulting EN-ADC(3) methods for IPs of molecules. Across the benchmarks considered here, EN-ADC(3) performs very well for closed-shell systems, reducing the mean absolute deviation from reference values relative to standard ADC(3) by a factor of almost four, and being almost twice as accurate as EOM-CCSD. Comprehensive amplitude renormalization by itself, as in the CCD-ADC(3) approach, does not lead to substantial improvements over ADC(3). It is rather the amplitude damping by the EN-shift that leads to the observed outstanding performance of EN-ADC(3).
Our data support the intuition outlined in the introduction that ADC(3) comes with the largest error for ionization on primarily electronegative atoms that leaves behind compact holes with large self-repulsion. This quantity is the main ingredient of the EN shift, whose magnitude therefore correlates strongly with the ADC(3) error and cancels most of its overestimation.

For open-shell systems, EN-ADC(3) halves the error of ADC(3) for weakly spin-contaminated molecules, outperforming EOM-CCSD, and still improves on ADC(3) for strongly spin-contaminated ones, for which EOM-CCSD remains more accurate. That the gains are smaller than for closed-shell molecules is readily rationalized by the same argument just given: 
Most of the systems contained in the open-shell set consist of radicals that leave behind a delocalized hole with small self-repulsion, which correlates with small ADC(3) errors and small EN shifts.

For series of linear acenes and oligothiophenes, the magnitude of the EN denominator shift remains sizable even for the largest systems considered. However, standard ADC(3) itself changes character along both series, moving from overestimating to underestimating the reference IPs.
This correlates with a decrease of the self-repulsion $J_{\text{HOMO}}$ of the ionized orbital as the hole delocalizes. The EN correction, which acts in a fixed direction and does not follow $J_{\text{HOMO}}$ as closely, improves agreement only where ADC(3) overestimates. Two series of this kind are insufficient to establish whether this behavior is general.

For practical applications, we recommend spin-summed ($2J-K$) EN-ADC(3) with the statically screened
singles shift and the $\Sigma(3+)$ static self-energy for closed-shell molecules.
It retains the accuracy of the best variant (MAD of 0.10 eV) and removes its only major outlier.
For open-shell systems, EN-ADC(3) with the $\Sigma(3)$ self-energy is recommended. 
Our data also suggest using the self-repulsion of the hole left behind after ionization as a simple
descriptor to gauge the reliability of an ADC(3) calculation.

Extensions of the present work to the ADC of the polarization propagator or to an ADC(4) of the self-energy suggest themselves. However, the simple amplitude substitutions leading to EN-ADC(3) are only valid for theories that do not already comprise the relevant 2p2h configurations. The diagonal pp ladders EN-dressing introduces are already present in the second-order doubles amplitudes that the ADC of the self-energy starting from fourth order\cite{Leitner2024Fourth-OrderMethods} and the ADC of the polarization propagator starting from third order require.\cite{Leitner2022ThePropagator} Naive EN variants of these approaches would count these interactions twice, and a double-counting correction would be required. An EN-ADC(2) of the polarization propagator would be free of double counting [by the same argument as CCD-ADC(2)\cite{Hodecker2019Algebraic-diagrammaticEnergies, Hodecker2019Algebraic-diagrammaticPolarizabilities}]. However, as it does not show a clear tendency to overestimate neutral excitation energies,\cite{Loos2020d} it is unlikely to offer clear benefits. The ADC(3) of the polarization propagator even tends to underestimate neutral excitation energies.\cite{Loos2020d} The ADC(4) of the self-energy also tends to underestimate charged excitations.\cite{Leitner2024Fourth-OrderMethods} The EN-shift damps the amplitudes and therefore reduces the energy of an excitation. For this reason, we do not consider such extensions to be worthwhile.

On the numerical side, EN partitioning carries practical limitations that its MP-based counterpart does not share. Reduced-scaling implementations of standard ADC(3) can be achieved through Laplace-type factorizations of the energy denominators,\cite{Almlof1991, Haser1992} which are unavailable once these denominators are dressed by the EN shift. 

On the theoretical side, our work suggests that renormalizing the ground-state amplitudes entering a self-energy is an interesting avenue toward improved approximations. Not surprisingly, the Green's function language naturally describes such renormalizations. We have shown that the EN-dressed amplitudes are bare vertices dressed by the diagonal elements of the time-ordered particle-particle propagator in the $T$-matrix approximation.\cite{Orlando2023} With THC- or ISDF-based factorizations, the full pp ladders could be calculated without raising the canonical $\mathcal{O}\left(N^5\right)$ scaling of ADC(3), which would be an interesting direction for future work.

Ladder-based renormalization is somewhat complementary to the vertex-based renormalization by \citet{Marie2026AnSelf-Energy}, who dress the outer scattering vertices of the ADC self-energy in the direct ph channel using the screened Coulomb interaction. EN partitioning instead dresses the scattering vertices in the pp-channel, using only the bare interaction already present in standard ADC(3). Making the connection between both approaches transparent would probably require calculating RPA screening in the Tamm--Dancoff approximation (TDA), in which ADC operates, and which makes it possible to express the effective Hamiltonians in terms of orbital energies and 4-center integrals only.\cite{Bintrim2021} Going beyond the TDA instead requires rotating the $U$ and $K+C$ blocks into the basis of RPA eigenvectors.\cite{Degroote2011, Bruneval2025GW+2SOSEXSemidefinite, Marie2026AnSelf-Energy} 

An interesting question would be whether pp ladders and such ph channel renormalizations can and should be combined without double counting, in the spirit of parquet approaches that treat multiple diagrammatic channels on an equal footing.\cite{DeDominicis1964, Bickers1989, Bickers2004, Rohringer2012, Marie2025ParquetApproximation}

\appendix
\section{\label{appendix}Connections to $T$-matrix Approximations}
In this appendix, we show that the pp-EN first-order doubles denominator coincides with the diagonal of the reducible time-ordered pp-propagators in the $T$-matrix approximation, and that the corresponding amplitudes are consequently given by bare scattering vertices dressed by these propagators. For an overview of the $T$-matrix approximation in a quantum chemistry context, we refer to Refs.~\citenum{Loos2022, Orlando2023, Marie2025AnomalousEquation, Marie2025ParquetApproximation, Marie2025Many-bodyChemistry}.

The first-order doubles amplitude of MPPT Eq.~\eqref{mp_amplitude2} might be expressed in terms of the irreducible time-ordered pair propagator
\begin{equation}
\label{eq:si-pair-propagator}
\begin{aligned}
L^{(0)}_{pq}(\omega)
 = &\frac{(1-n_p)(1-n_q)}{\omega - \epsilon_p - \epsilon_q + i\eta} \\
 & - \frac{n_p\,n_q}{\omega - \epsilon_p - \epsilon_q - i\eta} \;.
\end{aligned}
\end{equation}
Here, $n_p$ denotes the occupation of orbital $p$. For a virtual pair $(ab)$, $n_a = n_b = 0$, and only the forward branch of Eq.~\eqref{eq:si-pair-propagator} survives; for an occupied pair $(ij)$, $n_i = n_j = 1$, and only the backward branch survives. Introducing the short-hand
\begin{equation}
    D^{ab}_{ij} = \epsilon_i + \epsilon_j - \epsilon_a - \epsilon_b \;,
\end{equation}
Eq.~\eqref{mp_amplitude2} can be expressed as
\begin{equation}
\label{eq:si-2p2h-resolvent}
\begin{aligned}
&  t^{ab}_{ij} = \frac{\langle ab ||ij \rangle}{D^{ab}_{ij}} \\
 & =  
\lim_{\eta \rightarrow 0} \frac{i}{2\pi}\!\int\! d\omega\;
L^{(0)}_{ab}(\omega)\langle ab ||ij \rangle L^{(0)}_{ij}(\omega) \\
 &=  -\langle ab ||ij \rangle\lim_{\eta \rightarrow 0}  \frac{i}{2\pi} \\
 & \times \int\! d\omega\;
   \frac{1}{(\omega - \epsilon_a - \epsilon_b + i\eta)(\omega - \epsilon_i - \epsilon_j - i\eta)} \;.
 \end{aligned}
\end{equation}

We will now show that the EN doubles amplitudes can be expressed as
\begin{equation}
\label{eq:si-2p2h-resolvent_resummed}
\begin{aligned}
 \tilde{t}^{ab}_{ij} = & \frac{\langle ab ||ij \rangle}{D^{ab}_{ij} - \langle ij||ij \rangle - \langle ab||ab \rangle} \\
 = & 
\lim_{\eta \rightarrow 0} \frac{i}{2\pi}\!\int\! d\omega\;
L_{ab,ab}(\omega)\langle ab ||ij \rangle L_{ij,ij}(\omega) \;,
 \end{aligned}
\end{equation}
where $L$ is the diagonal of the reducible pair-propagator in the $T$-matrix approximation
\begin{equation}
\begin{aligned}
    L_{pq,pq}(\omega) = & L^{(0)}_{pq}(\omega) + L^{(0)}_{pq}(\omega)\langle pq\|rs\rangle L_{rs,pq}(\omega) \;.
\end{aligned}
\end{equation}
It is emphasized that the pair propagator remains a function of a single frequency because the bare interaction $\langle pq\|rs\rangle$ is instantaneous. Therefore, with Eq.~\eqref{eq:si-pair-propagator}, its diagonal may be resummed as an ordinary Dyson series
\begin{equation}
\label{eq:si-L-closed}
\begin{aligned}
L_{pq,pq}(\omega)
 = & \frac{L^{(0)}_{pq}(\omega)}{1-\langle pq\|pq\rangle\,L^{(0)}_{pq}(\omega)} \\
 = &\frac{(1-n_p)(1-n_q)}{\omega - \epsilon_p - \epsilon_q - \langle pq\|pq\rangle+ i\eta} \\
 & - \frac{n_p\,n_q}{\omega - \epsilon_p - \epsilon_q  + \langle pq\|pq\rangle - i\eta} \;.
 \end{aligned}
\end{equation}
Restricting to forward and backward branches, Eq.~\eqref{eq:si-L-closed} evaluates for the two pairs of interest we need in Eq.~\eqref{eq:si-2p2h-resolvent} to
\begin{equation}
\label{eq:si-L-explicit}
L_{ab,ab}(\omega) = \frac{1}{\omega-\epsilon_a-\epsilon_b-\langle ab||ab \rangle+i\eta}  \;,
\end{equation}
and 
\begin{equation}
\label{eq:si-L-explicit2}
L_{ij,ij}(\omega) = \frac{-1}{\omega-\epsilon_i-\epsilon_j+\langle ij||ij \rangle-i\eta} \;.
\end{equation}
Evaluating the frequency integral in Eq.~\eqref{eq:si-2p2h-resolvent_resummed} with \cref{eq:si-L-explicit,eq:si-L-explicit2}, standard residue calculus gives
\begin{equation}
\label{eq:si-resolvent-dressed}
\begin{aligned}
& i\!\int\!\frac{d\omega}{2\pi}\;
L_{ab,ab}(\omega)\,L_{ij,ij}(\omega) \\
 = &\frac{1}{D^{ab}_{ij} - \langle ij||ij \rangle - \langle ab||ab \rangle} \;,
 \end{aligned}
\end{equation}
which is precisely the EN-denominator restricted to the pp-channel of Eq.~\eqref{en_amplitude2}.

Expanding the denominators in powers of the diagonal interaction
shows that each order inserts one more diagonal rung of the pp ladder into the first-order doubles [Figure~\ref{fig:ladder_diagrams}(a)],
\begin{equation}
\label{eq:t-ladder-expansion}
\tilde{t}^{ab}_{ij} = t^{ab}_{ij} + t^{ab}_{ij}\,\frac{\langle ij||ij \rangle + \langle ab||ab \rangle}{D^{ab}_{ij}} + \dots \;.
\end{equation}
In the self-energy, the doubles amplitudes describe ground-state fluctuations, which enter through $U^{I/II,(2)}$ and $\Sigma(\infty)$.
Figure~\ref{fig:ladder_diagrams}(b) shows a third-order diagram of ADC(3) of this kind.
Inserting one rung gives the fourth-order diagrams (c) and (d).
Importantly, the rung acts on the 2p2h ground-state fluctuation before the external line enters, and not on the 2h1p state between the two outer vertices.
This shows that these diagrams are not generated by the resummation in $K+C$. They are therefore absent from ADC(3).

\begin{figure}[hbt!]
\centering
\tikzset{
 vx/.style={circle, fill=black, inner sep=1.1pt},
 vr/.style={circle, fill=red!70!black, inner sep=1.1pt},
 fl/.style={line width=0.8pt, postaction={decorate}, decoration={markings, mark=at position 0.5 with {\arrow{Stealth[length=5pt,width=4pt]}}}},
 int/.style={line width=0.8pt, dash pattern=on 3pt off 2pt},
 rint/.style={line width=0.9pt, dash pattern=on 3pt off 2pt, red!70!black},
 lb/.style={font=\footnotesize},
}
\begin{tikzpicture}[x=1cm,y=1cm]
\node at (-0.85,0.8) {$\tilde{t}^{ab}_{ij} \,=$};
\begin{scope}
 \draw[int] (0,0) -- (1,0); \node[vx] at (0,0) {}; \node[vx] at (1,0) {};
 \draw[fl] (0,0) -- (0,1.6) node[lb,above]{$a$};
 \draw[fl] (1,0) -- (1,1.6) node[lb,above]{$b$};
 \draw[fl] (-0.35,1.6) node[lb,above]{$i$} -- (0,0);
 \draw[fl] (1.35,1.6) node[lb,above]{$j$} -- (1,0);
\end{scope}
\node at (1.64,0.8) {$+$};
\begin{scope}[xshift=2.28cm]
 \draw[int] (0,0) -- (1,0); \node[vx] at (0,0) {}; \node[vx] at (1,0) {};
 \draw[fl] (0,0) -- (0,0.8); \draw[fl] (0,0.8) -- (0,1.6) node[lb,above]{$a$};
 \draw[fl] (1,0) -- (1,0.8); \draw[fl] (1,0.8) -- (1,1.6) node[lb,above]{$b$};
 \draw[rint] (0,0.8) -- (1,0.8); \node[vr] at (0,0.8) {}; \node[vr] at (1,0.8) {};
 \draw[fl] (-0.35,1.6) node[lb,above]{$i$} -- (0,0);
 \draw[fl] (1.35,1.6) node[lb,above]{$j$} -- (1,0);
\end{scope}
\node at (3.92,0.8) {$+$};
\begin{scope}[xshift=4.52cm]
 \draw[int] (0,0) -- (1,0); \node[vx] at (0,0) {}; \node[vx] at (1,0) {};
 \draw[fl] (0,1.6) node[lb,above]{$i$} -- (0,0.8); \draw[fl] (0,0.8) -- (0,0);
 \draw[fl] (1,1.6) node[lb,above]{$j$} -- (1,0.8); \draw[fl] (1,0.8) -- (1,0);
 \draw[rint] (0,0.8) -- (1,0.8); \node[vr] at (0,0.8) {}; \node[vr] at (1,0.8) {};
 \draw[fl] (0,0) -- (-0.35,1.6) node[lb,above]{$a$};
 \draw[fl] (1,0) -- (1.35,1.6) node[lb,above]{$b$};
\end{scope}
\node at (6.28,0.8) {$+\cdots$};
\node at (2.75,-0.6) {(a)};
\end{tikzpicture}

\vspace{0.6em}
\begin{tikzpicture}[x=1cm,y=1cm]
\foreach \sh/\rg/\pl in {0/0/(b),2.75/1/(c),5.5/2/(d)}{
\begin{scope}[xshift=\sh cm]
 \draw[int] (0,0) -- (1,0); \node[vx] at (0,0) {}; \node[vx] at (1,0) {};
 \ifnum\rg=1
  \draw[fl] (0,0) -- (0,0.75); \draw[fl] (0,0.75) -- (0,1.5);
  \draw[fl] (1,0) -- (1,0.75); \draw[fl] (1,0.75) -- (1,1.5);
  \draw[rint] (0,0.75) -- (1,0.75); \node[vr] at (0,0.75) {}; \node[vr] at (1,0.75) {};
 \else
  \draw[fl] (0,0) -- node[lb,right,pos=0.8]{$a$} (0,1.5);
  \draw[fl] (1,0) -- node[lb,left,pos=0.8]{$b$} (1,1.5);
 \fi
 \ifnum\rg=2
  \draw[fl] (-0.45,2.6) -- (-0.45,0.75); \draw[fl] (-0.45,0.75) to[out=-90,in=120] (0,0);
  \draw[fl] (1,2.6) to[out=-30,in=90] (1.45,2.0) -- (1.45,0.75); \draw[fl] (1.45,0.75) to[out=-90,in=60] (1,0);
  \draw[rint] (-0.45,0.75) -- (-0.08,0.75); \draw[rint] (0.08,0.75) -- (0.92,0.75); \draw[rint] (1.08,0.75) -- (1.45,0.75);
  \node[vr] at (-0.45,0.75) {}; \node[vr] at (1.45,0.75) {};
 \else
  \draw[fl] (-0.45,2.6) -- node[lb,left]{$r$} (-0.45,0.75) to[out=-90,in=120] (0,0);
  \draw[fl] (1,2.6) to[out=-30,in=90] (1.45,2.0) -- node[lb,right]{$s$} (1.45,0.75) to[out=-90,in=60] (1,0);
 \fi
 \draw[int] (0,1.5) -- (1,1.5); \node[vx] at (0,1.5) {}; \node[vx] at (1,1.5) {};
 \draw[line width=0.8pt] (0,1.5) -- (-0.25,1.8) node[lb,above]{$p$};
 \draw[fl] (1,1.5) -- node[lb,left]{$q$} (1,2.6);
 \draw[int] (-0.45,2.6) -- (1,2.6); \node[vx] at (-0.45,2.6) {}; \node[vx] at (1,2.6) {};
 \draw[line width=0.8pt] (-0.45,2.6) -- (-0.7,2.9) node[lb,above]{$p'$};
 \node at (0.5,-0.45) {\pl};
\end{scope}}
\end{tikzpicture}
\caption{
(a) Goldstone diagrams (time runs upward and dashed lines are antisymmetrized interactions)
for the expansion of the EN-dressed doubles amplitude, Eq.~\eqref{eq:t-ladder-expansion}, in powers of the diagonal interactions.
The red rungs are the diagonal elements $\langle ab\|ab\rangle$ (forward, pp) and $\langle ij\|ij\rangle$ (backward, hh) of the pp ladder.
(b-d) Third-order self-energy diagram (b) of ADC(3) from the term $-\tfrac{1}{2}\sum_{ab} t^{ab}_{rs}\langle pq\|ab\rangle$ of $U^{I,(2)}$, Eq.~\eqref{eq:u2_gen_I}, and fourth-order diagrams contained in EN-ADC(3) but not in ADC(3) obtained by inserting a pp or an hh rung into the ground-state fluctuation of (b).}
\label{fig:ladder_diagrams}
\end{figure}

\section{\label{app:factorization}Factorization of Time Orderings and Epstein--Nesbet Partitioning}
We here explain at second order why a consistent expansion of the self-energy in the EN partitioning does not take the ADC form with 2h1p and 2p1h intermediate states only.

Consider the removal part of the Green's function for an occupied orbital $p$,
\begin{equation}
    \label{eq:G_pp}
G^{-}_{pp}(\omega) = \braket{\Psi_0 | \hat{a}_p^{\dagger} \left[\omega - (E_0 - \hat{H})\right]^{-1} \hat{a}_p | \Psi_0} \;,
\end{equation}
and perturbatively expand the ground state,
\begin{equation}
    \ket{\Psi_0} = \ket{\Phi_0} + \sum_D t_D \ket{\Phi_D} + \dots
\end{equation}
and the resolvent in powers of the perturbation $\hat{V}$,
\begin{equation}
\left[\omega - (E_0 - \hat{H})\right]^{-1} = \hat{R}_0 - \hat{R}_0 \hat{V} \hat{R}_0 + \dots \;,
\end{equation}
with $\hat{R}_0 = \left[\omega - (E^{(0)}_0 - \hat{H}_0)\right]^{-1}$.
Let $D = (ij \to ab)$ with $p \notin \{i,j\}$ be a double excitation that produces the state
$\ket{\Phi_D} = \hat{a}_b^{\dagger}\hat{a}_a^{\dagger}\hat{a}_j\hat{a}_i\ket{\Phi_0}$ with energy $\varepsilon_D$.
Then, Eq.~\eqref{eq:G_pp} contains the 3h2p configuration
$\hat{a}_p\ket{\Phi_D}$ as an intermediate state,
whose zeroth-order energy relative to $\hat{a}_p\ket{\Phi_0}$ we write as
$\varepsilon_D + c_{pD}$. In case the double excitation is not affected by the propagation of the hole $p$, $c_{pD} = 0$.
To first order, the component of $\left[\omega - (E_0 - \hat{H})\right]^{-1} \hat{a}_p \ket{\Psi_0}$ along $\hat{a}_p\ket{\Phi_D}$
arises from two contributions that differ in the order in which the hole in $p$ and the ground state fluctuations are created.
If the fluctuation comes first, $\hat{a}_p$ acts on $t_D\ket{\Phi_D}$, and $\hat{R}_0$ propagates the 3h2p state.
If the hole comes first, $\hat{a}_p$ acts on $\ket{\Phi_0}$, $\hat{R}_0$ propagates $\hat{a}_p\ket{\Phi_0}$, $\hat{V}$ creates the fluctuation in the presence of the hole, and $\hat{R}_0$ propagates the 3h2p state.
Together, the two chains contribute the factor 
\begin{equation}
\label{eq:two-chains}
\begin{split}
& -\Braket{\Phi_D|\hat{V}|\Phi_0} \Bigg[ \frac{1}{\varepsilon_D\,(\omega - \epsilon_p + \varepsilon_D + c_{pD})} \\
& \quad + \frac{1}{(\omega - \epsilon_p)\,(\omega - \epsilon_p + \varepsilon_D + c_{pD})} \Bigg]
\end{split}
\end{equation}
to Eq.~\eqref{eq:G_pp}, i.e. a contribution due to the 3h2p/3p2h state.
If the hole propagates freely, while the ground state independently carries the fluctuation, $c_{pD} = 0$. 
In this case, the factorization theorem\cite{Shavitt2009Many-bodyTheory}
\begin{equation}
\label{eq:frantz-mills}
\begin{split}
& \frac{1}{\varepsilon_D\,(\omega - \epsilon_p + \varepsilon_D)} \\
& \quad + \frac{1}{(\omega - \epsilon_p)\,(\omega - \epsilon_p + \varepsilon_D)} = \frac{1}{\varepsilon_D\,(\omega - \epsilon_p)} \;,
\end{split}
\end{equation}
can be used to remove the poles at the 3h2p/3p2h configurations of Eq.~\eqref{eq:G_pp} due to Eq.~\eqref{eq:two-chains}. 
Equation~\eqref{eq:frantz-mills} is sometimes referred to as the Frantz--Mills theorem\cite{ai2025quantum}
and underlies the linked-diagram theorem.

In MP partitioning, zeroth-order energies are sums of orbital energies, and therefore $c_{pD} = 0$. 
As a result, the self-energy has no poles at 3h2p energies at this order.
The additivity of the MP zeroth-order energies is precisely the deficiency mentioned in the introduction, and the reason the EN shift has been introduced.
In MP theory, such poles first arise at fourth order from connected diagrams, which is why the 3h2p and 3p2h configurations enter the ADC of the self-energy only from ADC(4) onward.

In EN partitioning, the zeroth-order energy of $\hat{a}_p\ket{\Phi_D}$ contains the diagonal interactions between the additional hole and the quasiparticles of $\ket{\Phi_D}$,
\begin{equation}
c_{pD} = \langle pi||pi \rangle + \langle pj||pj \rangle - \langle pa||pa \rangle - \langle pb||pb \rangle \;,
\end{equation}
i.e. the diagonal interaction of the hole in $p$ with the two holes and two particles of $\ket{\Phi_D}$.
$c_{pD}$ connects the propagating hole to the ground-state fluctuation. 
For this reason, already at second order, a consistent EN-ADC has a pole at energy $\omega = \epsilon_p - \varepsilon_D - c_{pD}$,
and the corresponding supermatrix contains contributions due to the 3p2h/3h2p manifolds that would only arise starting from ADC(4) in MP partitioning.
\section*{Data and Software Availability Statement}
The data supporting the conclusions of this article are fully contained in the Supporting Information. All code is available at \href{https://github.com/ArnoFoerster/MBPTcode}{https://github.com/ArnoFoerster/MBPTcode}

\section*{Supporting Information}
All raw data calculated in this work, including the EN-ADC(3) results with the full EN denominators (ph terms included), the CCD-ADC(3) results, the descriptors of Table~\ref{tab:descriptors}, the $J_{\text{HOMO}}$ values underlying Figures~\ref{fig:adc3_error_vs_jhomo} and~\ref{fig:acenes_error_vs_jhomo}, a comparison table of CCSD(T) with IP-EOM-CCSDT IPs, and the optimized oligothiophene geometries.

\section*{Author Declarations} 
The author declares no conflict of interest.

\section*{Acknowledgments}
All code used to perform the calculations reported in this work, as well as the analysis and input scripts, has been largely written by Anthropic's Claude Code.
The author acknowledges financial support through a VENI grant from the Netherlands Organization for Scientific Research (NWO) under grant agreement VI.Veni.232.013.

\bibliography{references.bib,extra}

\providecommand{\latin}[1]{#1}
\makeatletter
\providecommand{\doi}
  {\begingroup\let\do\@makeother\dospecials
  \catcode`\{=1 \catcode`\}=2 \doi@aux}
\providecommand{\doi@aux}[1]{\endgroup\texttt{#1}}
\makeatother
\providecommand*\mcitethebibliography{\thebibliography}
\csname @ifundefined\endcsname{endmcitethebibliography}
  {\let\endmcitethebibliography\endthebibliography}{}
\begin{mcitethebibliography}{140}
\providecommand*\natexlab[1]{#1}
\providecommand*\mciteSetBstSublistMode[1]{}
\providecommand*\mciteSetBstMaxWidthForm[2]{}
\providecommand*\mciteBstWouldAddEndPuncttrue
  {\def\EndOfBibitem{\unskip.}}
\providecommand*\mciteBstWouldAddEndPunctfalse
  {\let\EndOfBibitem\relax}
\providecommand*\mciteSetBstMidEndSepPunct[3]{}
\providecommand*\mciteSetBstSublistLabelBeginEnd[3]{}
\providecommand*\EndOfBibitem{}
\mciteSetBstSublistMode{f}
\mciteSetBstMaxWidthForm{subitem}{(\alph{mcitesubitemcount})}
\mciteSetBstSublistLabelBeginEnd
  {\mcitemaxwidthsubitemform\space}
  {\relax}
  {\relax}

\bibitem[Schirmer \latin{et~al.}(1983)Schirmer, Cederbaum, and
  Walter]{Schirmer1983}
Schirmer,~J.; Cederbaum,~L.~S.; Walter,~O. {New approach to the one-particle
  Green's function for finite Fermi systems}. \emph{Physical Review A}
  \textbf{1983}, \emph{28}, 1237--1259\relax
\mciteBstWouldAddEndPuncttrue
\mciteSetBstMidEndSepPunct{\mcitedefaultmidpunct}
{\mcitedefaultendpunct}{\mcitedefaultseppunct}\relax
\EndOfBibitem
\bibitem[M{\o}ller and Plesset(1934)M{\o}ller, and Plesset]{Chr1934}
M{\o}ller,~C.; Plesset,~M.~S. {Note on an Approximation Treatment for
  Many-Electron Systems}. \emph{Physical Review} \textbf{1934}, \emph{46},
  618--622\relax
\mciteBstWouldAddEndPuncttrue
\mciteSetBstMidEndSepPunct{\mcitedefaultmidpunct}
{\mcitedefaultendpunct}{\mcitedefaultseppunct}\relax
\EndOfBibitem
\bibitem[Cederbaum(1975)]{Cederbaum1975}
Cederbaum,~L.~S. {Non-single-particle excitations in finite Fermi systems}.
  \emph{The Journal of Chemical Physics} \textbf{1975}, \emph{62},
  2160--2170\relax
\mciteBstWouldAddEndPuncttrue
\mciteSetBstMidEndSepPunct{\mcitedefaultmidpunct}
{\mcitedefaultendpunct}{\mcitedefaultseppunct}\relax
\EndOfBibitem
\bibitem[Hedin(1965)]{Hedin1965}
Hedin,~L. {New method for calculating the one-particle Green's function with
  application to the electron-gas problem}. \emph{Physical Review}
  \textbf{1965}, \emph{139}, A796\relax
\mciteBstWouldAddEndPuncttrue
\mciteSetBstMidEndSepPunct{\mcitedefaultmidpunct}
{\mcitedefaultendpunct}{\mcitedefaultseppunct}\relax
\EndOfBibitem
\bibitem[Stefanucci \latin{et~al.}(2014)Stefanucci, Pavlyukh, Uimonen, and van
  Leeuwen]{Stefanucci2014}
Stefanucci,~G.; Pavlyukh,~Y.; Uimonen,~A.~M.; van Leeuwen,~R. {Diagrammatic
  expansion for positive spectral functions beyond GW: Application to vertex
  corrections in the electron gas}. \emph{Physical Review B} \textbf{2014},
  \emph{90}, 115134\relax
\mciteBstWouldAddEndPuncttrue
\mciteSetBstMidEndSepPunct{\mcitedefaultmidpunct}
{\mcitedefaultendpunct}{\mcitedefaultseppunct}\relax
\EndOfBibitem
\bibitem[Bruneval \latin{et~al.}(2025)Bruneval, F{\"{o}}rster, and
  Pavlyukh]{Bruneval2025GW+2SOSEXSemidefinite}
Bruneval,~F.; F{\"{o}}rster,~A.; Pavlyukh,~Y. {GW+2SOSEX Self-Energy Made
  Positive Semidefinite}. \emph{Journal of Chemical Theory and Computation}
  \textbf{2025}, \emph{21}, 10223--10240\relax
\mciteBstWouldAddEndPuncttrue
\mciteSetBstMidEndSepPunct{\mcitedefaultmidpunct}
{\mcitedefaultendpunct}{\mcitedefaultseppunct}\relax
\EndOfBibitem
\bibitem[von Niessen \latin{et~al.}(1984)von Niessen, Schirmer, and
  Cederbaum]{vonNiessen1984ComputationalFunction}
von Niessen,~W.; Schirmer,~J.; Cederbaum,~L.~S. {Computational methods for the
  one-particle green's function}. \emph{Computer Physics Reports}
  \textbf{1984}, \emph{1}, 57--125\relax
\mciteBstWouldAddEndPuncttrue
\mciteSetBstMidEndSepPunct{\mcitedefaultmidpunct}
{\mcitedefaultendpunct}{\mcitedefaultseppunct}\relax
\EndOfBibitem
\bibitem[Winter(1972)]{Winter1972StudyFunction}
Winter,~J. {Study of core excitations in one-particle and one-hole nuclei by
  means of the six-point Green function}. \emph{Nuclear Physics A}
  \textbf{1972}, \emph{194}, 535--551\relax
\mciteBstWouldAddEndPuncttrue
\mciteSetBstMidEndSepPunct{\mcitedefaultmidpunct}
{\mcitedefaultendpunct}{\mcitedefaultseppunct}\relax
\EndOfBibitem
\bibitem[Banerjee and Sokolov(2021)Banerjee, and
  Sokolov]{Banerjee2021EfficientPairs}
Banerjee,~S.; Sokolov,~A.~Y. {Efficient implementation of the single-reference
  algebraic diagrammatic construction theory for charged excitations:
  Applications to the TEMPO radical and DNA base pairs}. \emph{Journal of
  Chemical Physics} \textbf{2021}, \emph{154}, 074105\relax
\mciteBstWouldAddEndPuncttrue
\mciteSetBstMidEndSepPunct{\mcitedefaultmidpunct}
{\mcitedefaultendpunct}{\mcitedefaultseppunct}\relax
\EndOfBibitem
\bibitem[Dolgounitcheva \latin{et~al.}(2016)Dolgounitcheva, D{\'{i}}az-Tinoco,
  Zakrzewski, Richard, Marom, Sherrill, and Ortiz]{Dolgounitcheva2016}
Dolgounitcheva,~O.; D{\'{i}}az-Tinoco,~M.; Zakrzewski,~V.~G.; Richard,~R.~M.;
  Marom,~N.; Sherrill,~C.~D.; Ortiz,~J.~V. {Accurate Ionization Potentials and
  Electron Affinities of Acceptor Molecules IV: Electron-Propagator Methods}.
  \emph{Journal of Chemical Theory and Computation} \textbf{2016}, \emph{12},
  627--637\relax
\mciteBstWouldAddEndPuncttrue
\mciteSetBstMidEndSepPunct{\mcitedefaultmidpunct}
{\mcitedefaultendpunct}{\mcitedefaultseppunct}\relax
\EndOfBibitem
\bibitem[Opoku \latin{et~al.}(2023)Opoku, Paw{\l}owski, and Ortiz]{Opoku2023}
Opoku,~E.; Paw{\l}owski,~F.; Ortiz,~J.~V. {A new generation of non-diagonal,
  renormalized self-energies for calculation of electron removal energies}.
  \emph{Journal of Chemical Physics} \textbf{2023}, \emph{159}, 124109\relax
\mciteBstWouldAddEndPuncttrue
\mciteSetBstMidEndSepPunct{\mcitedefaultmidpunct}
{\mcitedefaultendpunct}{\mcitedefaultseppunct}\relax
\EndOfBibitem
\bibitem[Banerjee and Sokolov(2023)Banerjee, and
  Sokolov]{Banerjee2023AlgebraicSpectra}
Banerjee,~S.; Sokolov,~A.~Y. {Algebraic Diagrammatic Construction Theory for
  Simulating Charged Excited States and Photoelectron Spectra}. \emph{Journal
  of Chemical Theory and Computation} \textbf{2023}, \emph{19},
  3037--3053\relax
\mciteBstWouldAddEndPuncttrue
\mciteSetBstMidEndSepPunct{\mcitedefaultmidpunct}
{\mcitedefaultendpunct}{\mcitedefaultseppunct}\relax
\EndOfBibitem
\bibitem[Sokolov(2018)]{Sokolov2018Multi-referenceImplementation}
Sokolov,~A.~Y. {Multi-reference algebraic diagrammatic construction theory for
  excited states: General formulation and first-order implementation}.
  \emph{Journal of Chemical Physics} \textbf{2018}, \emph{149}, 204113\relax
\mciteBstWouldAddEndPuncttrue
\mciteSetBstMidEndSepPunct{\mcitedefaultmidpunct}
{\mcitedefaultendpunct}{\mcitedefaultseppunct}\relax
\EndOfBibitem
\bibitem[Chatterjee and Sokolov(2019)Chatterjee, and
  Sokolov]{Chatterjee2019Second-OrderSystems}
Chatterjee,~K.; Sokolov,~A.~Y. {Second-Order Multireference Algebraic
  Diagrammatic Construction Theory for Photoelectron Spectra of Strongly
  Correlated Systems}. \emph{Journal of Chemical Theory and Computation}
  \textbf{2019}, \emph{15}, 5908--5924\relax
\mciteBstWouldAddEndPuncttrue
\mciteSetBstMidEndSepPunct{\mcitedefaultmidpunct}
{\mcitedefaultendpunct}{\mcitedefaultseppunct}\relax
\EndOfBibitem
\bibitem[Chatterjee and Sokolov(2020)Chatterjee, and
  Sokolov]{Chatterjee2020ExtendedExcitations}
Chatterjee,~K.; Sokolov,~A.~Y. {Extended Second-Order Multireference Algebraic
  Diagrammatic Construction Theory for Charged Excitations}. \emph{Journal of
  Chemical Theory and Computation} \textbf{2020}, \emph{16}, 6343--6357\relax
\mciteBstWouldAddEndPuncttrue
\mciteSetBstMidEndSepPunct{\mcitedefaultmidpunct}
{\mcitedefaultendpunct}{\mcitedefaultseppunct}\relax
\EndOfBibitem
\bibitem[Mazin and Sokolov(2021)Mazin, and
  Sokolov]{Mazin2021MultireferenceBenchmark}
Mazin,~I.~M.; Sokolov,~A.~Y. {Multireference Algebraic Diagrammatic
  Construction Theory for Excited States: Extended Second-Order Implementation
  and Benchmark}. \emph{Journal of Chemical Theory and Computation}
  \textbf{2021}, \emph{17}, 6152--6165\relax
\mciteBstWouldAddEndPuncttrue
\mciteSetBstMidEndSepPunct{\mcitedefaultmidpunct}
{\mcitedefaultendpunct}{\mcitedefaultseppunct}\relax
\EndOfBibitem
\bibitem[De~Moura and Sokolov(2022)De~Moura, and
  Sokolov]{DeMoura2022SimulatingTheory}
De~Moura,~C.~E.; Sokolov,~A.~Y. {Simulating X-ray photoelectron spectra with
  strong electron correlation using multireference algebraic diagrammatic
  construction theory}. \emph{Physical Chemistry Chemical Physics}
  \textbf{2022}, \emph{24}, 4769--4784\relax
\mciteBstWouldAddEndPuncttrue
\mciteSetBstMidEndSepPunct{\mcitedefaultmidpunct}
{\mcitedefaultendpunct}{\mcitedefaultseppunct}\relax
\EndOfBibitem
\bibitem[Olsen \latin{et~al.}(1996)Olsen, Christiansen, Koch, and
  J{\o}rgensen]{Olsen1996SurprisingTheory}
Olsen,~J.; Christiansen,~O.; Koch,~H.; J{\o}rgensen,~P. {Surprising cases of
  divergent behavior in M{\o}ller-Plesset perturbation theory}. \emph{Journal
  of Chemical Physics} \textbf{1996}, \emph{105}, 5082--5090\relax
\mciteBstWouldAddEndPuncttrue
\mciteSetBstMidEndSepPunct{\mcitedefaultmidpunct}
{\mcitedefaultendpunct}{\mcitedefaultseppunct}\relax
\EndOfBibitem
\bibitem[Hirata \latin{et~al.}(2024)Hirata, Grabowski, Ortiz, and
  Bartlett]{Hirata2024a}
Hirata,~S.; Grabowski,~I.; Ortiz,~J.~V.; Bartlett,~R.~J. {Nonconvergence of the
  Feynman-Dyson diagrammatic perturbation expansion of propagators}.
  \emph{Physical Review A} \textbf{2024}, \emph{109}, 052220\relax
\mciteBstWouldAddEndPuncttrue
\mciteSetBstMidEndSepPunct{\mcitedefaultmidpunct}
{\mcitedefaultendpunct}{\mcitedefaultseppunct}\relax
\EndOfBibitem
\bibitem[Raghavachari \latin{et~al.}(1989)Raghavachari, Trucks, Pople, and
  Head-Gordon]{Raghavachari1989}
Raghavachari,~K.; Trucks,~G.; Pople,~J.~A.; Head-Gordon,~M. {A fifth-order
  perturbation comparison of electron correlation theories}. \emph{Chemical
  Physics Letters} \textbf{1989}, \emph{157}, 479--483\relax
\mciteBstWouldAddEndPuncttrue
\mciteSetBstMidEndSepPunct{\mcitedefaultmidpunct}
{\mcitedefaultendpunct}{\mcitedefaultseppunct}\relax
\EndOfBibitem
\bibitem[Stanton(1997)]{stanton1997ccsd}
Stanton,~J.~F. Why CCSD (T) works: a different perspective. \emph{Chemical
  Physics Letters} \textbf{1997}, \emph{281}, 130--134\relax
\mciteBstWouldAddEndPuncttrue
\mciteSetBstMidEndSepPunct{\mcitedefaultmidpunct}
{\mcitedefaultendpunct}{\mcitedefaultseppunct}\relax
\EndOfBibitem
\bibitem[Nguyen \latin{et~al.}(2020)Nguyen, Chen, Agee, Burow, Tang, and
  Furche]{Nguyen2020}
Nguyen,~B.~D.; Chen,~G.~P.; Agee,~M.~M.; Burow,~A.~M.; Tang,~M.~P.; Furche,~F.
  {Divergence of Many-Body Perturbation Theory for Noncovalent Interactions of
  Large Molecules}. \emph{Journal of Chemical Theory and Computation}
  \textbf{2020}, \emph{16}, 2258--2273\relax
\mciteBstWouldAddEndPuncttrue
\mciteSetBstMidEndSepPunct{\mcitedefaultmidpunct}
{\mcitedefaultendpunct}{\mcitedefaultseppunct}\relax
\EndOfBibitem
\bibitem[Al-Hamdani \latin{et~al.}(2021)Al-Hamdani, Nagy, Zen, Barton,
  K{\'{a}}llay, Brandenburg, and Tkatchenko]{Al-Hamdani2021}
Al-Hamdani,~Y.~S.; Nagy,~P.~R.; Zen,~A.; Barton,~D.; K{\'{a}}llay,~M.;
  Brandenburg,~J.~G.; Tkatchenko,~A. {Interactions between large molecules pose
  a puzzle for reference quantum mechanical methods}. \emph{Nature
  Communications} \textbf{2021}, \emph{12}, 3927\relax
\mciteBstWouldAddEndPuncttrue
\mciteSetBstMidEndSepPunct{\mcitedefaultmidpunct}
{\mcitedefaultendpunct}{\mcitedefaultseppunct}\relax
\EndOfBibitem
\bibitem[Sch{\"a}fer \latin{et~al.}(2025)Sch{\"a}fer, Irmler, Gallo, and
  Gr{\"u}neis]{schafer2025understanding}
Sch{\"a}fer,~T.; Irmler,~A.; Gallo,~A.; Gr{\"u}neis,~A. Understanding
  discrepancies in noncovalent interaction energies from wavefunction theories
  for large molecules. \emph{Nature Communications} \textbf{2025}, \emph{16},
  9108\relax
\mciteBstWouldAddEndPuncttrue
\mciteSetBstMidEndSepPunct{\mcitedefaultmidpunct}
{\mcitedefaultendpunct}{\mcitedefaultseppunct}\relax
\EndOfBibitem
\bibitem[Bartlett and Musia{\l}(2007)Bartlett, and
  Musia{\l}]{Bartlett2007Coupled-clusterChemistry}
Bartlett,~R.~J.; Musia{\l},~M. {Coupled-cluster theory in quantum chemistry}.
  \emph{Reviews of Modern Physics} \textbf{2007}, \emph{79}, 291--352\relax
\mciteBstWouldAddEndPuncttrue
\mciteSetBstMidEndSepPunct{\mcitedefaultmidpunct}
{\mcitedefaultendpunct}{\mcitedefaultseppunct}\relax
\EndOfBibitem
\bibitem[Fink(2016)]{Fink2016}
Fink,~R.~F. {Why does MP2 work?} \emph{Journal of Chemical Physics}
  \textbf{2016}, \emph{145}, 184101\relax
\mciteBstWouldAddEndPuncttrue
\mciteSetBstMidEndSepPunct{\mcitedefaultmidpunct}
{\mcitedefaultendpunct}{\mcitedefaultseppunct}\relax
\EndOfBibitem
\bibitem[Masios \latin{et~al.}(2023)Masios, Irmler, Sch{\"{a}}fer, and
  Gr{\"{u}}neis]{Masios2023}
Masios,~N.; Irmler,~A.; Sch{\"{a}}fer,~T.; Gr{\"{u}}neis,~A. {Averting the
  Infrared Catastrophe in the Gold Standard of Quantum Chemistry}.
  \emph{Physical Review Letters} \textbf{2023}, \emph{131}, 186401\relax
\mciteBstWouldAddEndPuncttrue
\mciteSetBstMidEndSepPunct{\mcitedefaultmidpunct}
{\mcitedefaultendpunct}{\mcitedefaultseppunct}\relax
\EndOfBibitem
\bibitem[Bruneval and F{\"{o}}rster(2024)Bruneval, and
  F{\"{o}}rster]{Bruneval2024}
Bruneval,~F.; F{\"{o}}rster,~A. {Fully dynamic G3W2 self-energy for finite
  systems: Formulas and benchmark}. \emph{Journal of Chemical Theory and
  Computation} \textbf{2024}, \emph{20}, 3218−3230\relax
\mciteBstWouldAddEndPuncttrue
\mciteSetBstMidEndSepPunct{\mcitedefaultmidpunct}
{\mcitedefaultendpunct}{\mcitedefaultseppunct}\relax
\EndOfBibitem
\bibitem[Marie \latin{et~al.}(2026)Marie, T{\"{o}}lle, and
  Loos]{Marie2026AnSelf-Energy}
Marie,~A.; T{\"{o}}lle,~J.; Loos,~P.-F. {An Algebraic-Diagrammatic Construction
  for Vertex Corrections to the GW Self-Energy}. \emph{The Journal of Physical
  Chemistry Letters} \textbf{2026}, \emph{17}, 8541--8551\relax
\mciteBstWouldAddEndPuncttrue
\mciteSetBstMidEndSepPunct{\mcitedefaultmidpunct}
{\mcitedefaultendpunct}{\mcitedefaultseppunct}\relax
\EndOfBibitem
\bibitem[Macke(1950)]{Macke1950}
Macke,~W. {{\"{U}}ber die Wechselwirkungen im Fermi-Gas}. \emph{Zeitschrift
  f{\"{u}}r Naturforschung} \textbf{1950}, \emph{5}, 192--208\relax
\mciteBstWouldAddEndPuncttrue
\mciteSetBstMidEndSepPunct{\mcitedefaultmidpunct}
{\mcitedefaultendpunct}{\mcitedefaultseppunct}\relax
\EndOfBibitem
\bibitem[Gell-Mann and Brueckner(1957)Gell-Mann, and Brueckner]{Gell-Mann1957}
Gell-Mann,~M.; Brueckner,~K.~A. {Correlation energy of an electron gas at high
  density}. \emph{Physical Review} \textbf{1957}, \emph{106}, 364--368\relax
\mciteBstWouldAddEndPuncttrue
\mciteSetBstMidEndSepPunct{\mcitedefaultmidpunct}
{\mcitedefaultendpunct}{\mcitedefaultseppunct}\relax
\EndOfBibitem
\bibitem[Epstein(1926)]{epstein1926stark}
Epstein,~P.~S. The Stark effect from the point of view of Schroedinger's
  quantum theory. \emph{Physical Review} \textbf{1926}, \emph{28}, 695\relax
\mciteBstWouldAddEndPuncttrue
\mciteSetBstMidEndSepPunct{\mcitedefaultmidpunct}
{\mcitedefaultendpunct}{\mcitedefaultseppunct}\relax
\EndOfBibitem
\bibitem[Nesbet(1955)]{nesbet1955configuration}
Nesbet,~R. Configuration interaction in orbital theories. \emph{Proceedings of
  the Royal Society of London. Series A. Mathematical and Physical Sciences}
  \textbf{1955}, \emph{230}, 312--321\relax
\mciteBstWouldAddEndPuncttrue
\mciteSetBstMidEndSepPunct{\mcitedefaultmidpunct}
{\mcitedefaultendpunct}{\mcitedefaultseppunct}\relax
\EndOfBibitem
\bibitem[Shavitt and Bartlett(2009)Shavitt, and
  Bartlett]{Shavitt2009Many-bodyTheory}
Shavitt,~I.; Bartlett,~R.~J. \emph{{Many-body methods in chemistry and physics:
  MBPT and coupled-cluster theory}}; Cambridge University Press, 2009\relax
\mciteBstWouldAddEndPuncttrue
\mciteSetBstMidEndSepPunct{\mcitedefaultmidpunct}
{\mcitedefaultendpunct}{\mcitedefaultseppunct}\relax
\EndOfBibitem
\bibitem[Jiang and Engel(2006)Jiang, and Engel]{Jiang2006}
Jiang,~H.; Engel,~E. {Kohn-Sham perturbation theory: Simple solution to
  variational instability of second order correlation energy functional}.
  \emph{Journal of Chemical Physics} \textbf{2006}, \emph{125}, 184108\relax
\mciteBstWouldAddEndPuncttrue
\mciteSetBstMidEndSepPunct{\mcitedefaultmidpunct}
{\mcitedefaultendpunct}{\mcitedefaultseppunct}\relax
\EndOfBibitem
\bibitem[Jiang and Engel(2007)Jiang, and Engel]{Jiang2007}
Jiang,~H.; Engel,~E. {Random-phase-approximation-based correlation energy
  functionals: Benchmark results for atoms}. \emph{Journal of Chemical Physics}
  \textbf{2007}, \emph{127}, 184108\relax
\mciteBstWouldAddEndPuncttrue
\mciteSetBstMidEndSepPunct{\mcitedefaultmidpunct}
{\mcitedefaultendpunct}{\mcitedefaultseppunct}\relax
\EndOfBibitem
\bibitem[Szabo and Ostlund(2012)Szabo, and Ostlund]{Szabo2012}
Szabo,~A.; Ostlund,~N.~S. \emph{{Modern quantum chemistry: introduction to
  advanced electronic structure theory}}; Dover Publications INC. New York,
  2012\relax
\mciteBstWouldAddEndPuncttrue
\mciteSetBstMidEndSepPunct{\mcitedefaultmidpunct}
{\mcitedefaultendpunct}{\mcitedefaultseppunct}\relax
\EndOfBibitem
\bibitem[Orlando \latin{et~al.}(2023)Orlando, Romaniello, and
  Loos]{Orlando2023}
Orlando,~R.; Romaniello,~P.; Loos,~P.~F. {The three channels of many-body
  perturbation theory: GW, particle-particle, and electron-hole T-matrix
  self-energies}. \emph{Journal of Chemical Physics} \textbf{2023}, \emph{159},
  184113\relax
\mciteBstWouldAddEndPuncttrue
\mciteSetBstMidEndSepPunct{\mcitedefaultmidpunct}
{\mcitedefaultendpunct}{\mcitedefaultseppunct}\relax
\EndOfBibitem
\bibitem[Marie \latin{et~al.}(2025)Marie, Romaniello, Blase, and
  Loos]{Marie2025AnomalousEquation}
Marie,~A.; Romaniello,~P.; Blase,~X.; Loos,~P.~F. {Anomalous propagators and
  the particle-particle channel: Bethe-Salpeter equation}. \emph{Journal of
  Chemical Physics} \textbf{2025}, \emph{162}, 134105\relax
\mciteBstWouldAddEndPuncttrue
\mciteSetBstMidEndSepPunct{\mcitedefaultmidpunct}
{\mcitedefaultendpunct}{\mcitedefaultseppunct}\relax
\EndOfBibitem
\bibitem[Marie and Loos(2025)Marie, and Loos]{Marie2025ParquetApproximation}
Marie,~A.; Loos,~P.-F. {Parquet theory for molecular systems: Formalism and
  static kernel parquet approximation}. \emph{The Journal of Chemical Physics}
  \textbf{2025}, \emph{163}, 194115\relax
\mciteBstWouldAddEndPuncttrue
\mciteSetBstMidEndSepPunct{\mcitedefaultmidpunct}
{\mcitedefaultendpunct}{\mcitedefaultseppunct}\relax
\EndOfBibitem
\bibitem[Marie(2025)]{Marie2025Many-bodyChemistry}
Marie,~A. {Many-body perturbation theory in quantum chemistry}. Ph.D.\ thesis,
  2025\relax
\mciteBstWouldAddEndPuncttrue
\mciteSetBstMidEndSepPunct{\mcitedefaultmidpunct}
{\mcitedefaultendpunct}{\mcitedefaultseppunct}\relax
\EndOfBibitem
\bibitem[Fink(2006)]{Fink2006TwoEnergy}
Fink,~R.~F. {Two new unitary-invariant and size-consistent perturbation
  theoretical approaches to the electron correlation energy}. \emph{Chemical
  Physics Letters} \textbf{2006}, \emph{428}, 461--466\relax
\mciteBstWouldAddEndPuncttrue
\mciteSetBstMidEndSepPunct{\mcitedefaultmidpunct}
{\mcitedefaultendpunct}{\mcitedefaultseppunct}\relax
\EndOfBibitem
\bibitem[Leitner \latin{et~al.}(2026)Leitner, Dittmer, Schneider, Behnle, Fink,
  and Dreuw]{Leitner2026RE-ADC:Partitioning}
Leitner,~J.; Dittmer,~L.~B.; Schneider,~F.; Behnle,~S.; Fink,~R.~F.; Dreuw,~A.
  {RE-ADC: The algebraic diagrammatic construction scheme for the polarization
  propagator using the retaining-the-excitation-degree partitioning}.
  \emph{Journal of Chemical Physics} \textbf{2026}, \emph{164}, 164101\relax
\mciteBstWouldAddEndPuncttrue
\mciteSetBstMidEndSepPunct{\mcitedefaultmidpunct}
{\mcitedefaultendpunct}{\mcitedefaultseppunct}\relax
\EndOfBibitem
\bibitem[Keller \latin{et~al.}(2022)Keller, Tsatsoulis, Reuter, and
  Margraf]{Keller2022}
Keller,~E.; Tsatsoulis,~T.; Reuter,~K.; Margraf,~J.~T. {Regularized
  second-order correlation methods for extended systems}. \emph{Journal of
  Chemical Physics} \textbf{2022}, \emph{156}, 024106\relax
\mciteBstWouldAddEndPuncttrue
\mciteSetBstMidEndSepPunct{\mcitedefaultmidpunct}
{\mcitedefaultendpunct}{\mcitedefaultseppunct}\relax
\EndOfBibitem
\bibitem[Carter-Fenk and Head-Gordon(2023)Carter-Fenk, and
  Head-Gordon]{Carter-Fenk2023RepartitionedEnergy}
Carter-Fenk,~K.; Head-Gordon,~M. {Repartitioned Brillouin-Wigner perturbation
  theory with a size-consistent second-order correlation energy}. \emph{Journal
  of Chemical Physics} \textbf{2023}, \emph{158}, 234108\relax
\mciteBstWouldAddEndPuncttrue
\mciteSetBstMidEndSepPunct{\mcitedefaultmidpunct}
{\mcitedefaultendpunct}{\mcitedefaultseppunct}\relax
\EndOfBibitem
\bibitem[Dittmer and Head-Gordon(2025)Dittmer, and
  Head-Gordon]{Dittmer2025RepartitioningModels}
Dittmer,~L.~B.; Head-Gordon,~M. {Repartitioning the Hamiltonian in many-body
  second-order Brillouin-Wigner perturbation theory: Uncovering new
  size-consistent models}. \emph{Journal of Chemical Physics} \textbf{2025},
  \emph{162}, 054109\relax
\mciteBstWouldAddEndPuncttrue
\mciteSetBstMidEndSepPunct{\mcitedefaultmidpunct}
{\mcitedefaultendpunct}{\mcitedefaultseppunct}\relax
\EndOfBibitem
\bibitem[Dittmer \latin{et~al.}(2026)Dittmer, M{\"{u}}ller, Leitner, Dempwolff,
  and Dreuw]{Dittmer2026BWs-ADC:Theory}
Dittmer,~L.~B.; M{\"{u}}ller,~A.~J.; Leitner,~J.; Dempwolff,~A.~L.; Dreuw,~A.
  {BWs-ADC: The algebraic diagrammatic construction for the polarization
  propagator using a size-consistent Brillouin-Wigner partitioning up to third
  order in perturbation theory}. \emph{Journal of Chemical Physics}
  \textbf{2026}, \emph{165}, 094105\relax
\mciteBstWouldAddEndPuncttrue
\mciteSetBstMidEndSepPunct{\mcitedefaultmidpunct}
{\mcitedefaultendpunct}{\mcitedefaultseppunct}\relax
\EndOfBibitem
\bibitem[{\v{C}}{\'{i}}{\v{z}}ek(1966)]{Cizek1966}
{\v{C}}{\'{i}}{\v{z}}ek,~J. {On the Correlation Problem in Atomic and Molecular
  Systems. Calculation of Wavefunction Components in Ursell-Type Expansion
  Using Quantum-Field Theoretical Methods}. \emph{The Journal of Chemical
  Physics} \textbf{1966}, \emph{45}, 4256--4266\relax
\mciteBstWouldAddEndPuncttrue
\mciteSetBstMidEndSepPunct{\mcitedefaultmidpunct}
{\mcitedefaultendpunct}{\mcitedefaultseppunct}\relax
\EndOfBibitem
\bibitem[Taube and Bartlett(2009)Taube, and Bartlett]{taube2009rethinking}
Taube,~A.~G.; Bartlett,~R.~J. Rethinking linearized coupled-cluster theory.
  \emph{The Journal of chemical physics} \textbf{2009}, \emph{130},
  144112\relax
\mciteBstWouldAddEndPuncttrue
\mciteSetBstMidEndSepPunct{\mcitedefaultmidpunct}
{\mcitedefaultendpunct}{\mcitedefaultseppunct}\relax
\EndOfBibitem
\bibitem[Roos and Andersson(1995)Roos, and
  Andersson]{roos1995multiconfigurational}
Roos,~B.~O.; Andersson,~K. Multiconfigurational perturbation theory with level
  shift—the Cr2 potential revisited. \emph{Chemical physics letters}
  \textbf{1995}, \emph{245}, 215--223\relax
\mciteBstWouldAddEndPuncttrue
\mciteSetBstMidEndSepPunct{\mcitedefaultmidpunct}
{\mcitedefaultendpunct}{\mcitedefaultseppunct}\relax
\EndOfBibitem
\bibitem[St{\"u}ck and Head-Gordon(2013)St{\"u}ck, and
  Head-Gordon]{stuck2013regularized}
St{\"u}ck,~D.; Head-Gordon,~M. Regularized orbital-optimized second-order
  perturbation theory. \emph{The Journal of chemical physics} \textbf{2013},
  \emph{139}, 244109\relax
\mciteBstWouldAddEndPuncttrue
\mciteSetBstMidEndSepPunct{\mcitedefaultmidpunct}
{\mcitedefaultendpunct}{\mcitedefaultseppunct}\relax
\EndOfBibitem
\bibitem[Lee and Head-Gordon(2018)Lee, and Head-Gordon]{Lee2018}
Lee,~J.; Head-Gordon,~M. {Regularized Orbital-Optimized Second-Order
  M{\o}ller-Plesset Perturbation Theory: A Reliable Fifth-Order-Scaling
  Electron Correlation Model with Orbital Energy Dependent Regularizers}.
  \emph{Journal of Chemical Theory and Computation} \textbf{2018}, \emph{14},
  5203--5219\relax
\mciteBstWouldAddEndPuncttrue
\mciteSetBstMidEndSepPunct{\mcitedefaultmidpunct}
{\mcitedefaultendpunct}{\mcitedefaultseppunct}\relax
\EndOfBibitem
\bibitem[Marie and Loos(2023)Marie, and Loos]{Marie2023a}
Marie,~A.; Loos,~P.-F. {A Similarity Renormalization Group Approach to
  Green’s Function Methods}. \emph{Journal of Chemical Theory and
  Computation} \textbf{2023}, \emph{19}, 3943−3957\relax
\mciteBstWouldAddEndPuncttrue
\mciteSetBstMidEndSepPunct{\mcitedefaultmidpunct}
{\mcitedefaultendpunct}{\mcitedefaultseppunct}\relax
\EndOfBibitem
\bibitem[Krieger and T{\"{o}}lle(2026)Krieger, and
  T{\"{o}}lle]{Krieger2026RenormalizationTheory}
Krieger,~J.; T{\"{o}}lle,~J. {Renormalization group approach to second-order
  Green’s function theory}. \emph{The Journal of Chemical Physics}
  \textbf{2026}, \emph{164}, 174106\relax
\mciteBstWouldAddEndPuncttrue
\mciteSetBstMidEndSepPunct{\mcitedefaultmidpunct}
{\mcitedefaultendpunct}{\mcitedefaultseppunct}\relax
\EndOfBibitem
\bibitem[Leininger \latin{et~al.}(2000)Leininger, Allen, Schaefer~III, and
  Sherrill]{leininger2000mo}
Leininger,~M.~L.; Allen,~W.~D.; Schaefer~III,~H.~F.; Sherrill,~C.~D. Is
  Møller--Plesset perturbation theory a convergent ab initio method? \emph{The
  Journal of Chemical Physics} \textbf{2000}, \emph{112}, 9213--9222\relax
\mciteBstWouldAddEndPuncttrue
\mciteSetBstMidEndSepPunct{\mcitedefaultmidpunct}
{\mcitedefaultendpunct}{\mcitedefaultseppunct}\relax
\EndOfBibitem
\bibitem[Anderson(1958)]{anderson1958random}
Anderson,~P.~W. Random-phase approximation in the theory of superconductivity.
  \emph{Physical Review} \textbf{1958}, \emph{112}, 1900\relax
\mciteBstWouldAddEndPuncttrue
\mciteSetBstMidEndSepPunct{\mcitedefaultmidpunct}
{\mcitedefaultendpunct}{\mcitedefaultseppunct}\relax
\EndOfBibitem
\bibitem[Baym and Kadanoff(1961)Baym, and Kadanoff]{Baym1961}
Baym,~G.; Kadanoff,~L.~P. {Conservation laws and correlation functions}.
  \emph{Physical Review} \textbf{1961}, \emph{124}, 287--299\relax
\mciteBstWouldAddEndPuncttrue
\mciteSetBstMidEndSepPunct{\mcitedefaultmidpunct}
{\mcitedefaultendpunct}{\mcitedefaultseppunct}\relax
\EndOfBibitem
\bibitem[Loos and Romaniello(2022)Loos, and Romaniello]{Loos2022}
Loos,~P.~F.; Romaniello,~P. {Static and dynamic Bethe-Salpeter equations in the
  T-matrix approximation}. \emph{Journal of Chemical Physics} \textbf{2022},
  \emph{156}, 164101\relax
\mciteBstWouldAddEndPuncttrue
\mciteSetBstMidEndSepPunct{\mcitedefaultmidpunct}
{\mcitedefaultendpunct}{\mcitedefaultseppunct}\relax
\EndOfBibitem
\bibitem[Bethe and Goldstone(1957)Bethe, and Goldstone]{bethe1957effect}
Bethe,~H.~A.; Goldstone,~J. Effect of a repulsive core in the theory of complex
  nuclei. \emph{Proceedings of the Royal Society of London. Series A,
  Mathematical and Physical Sciences} \textbf{1957}, \emph{238}, 551--567\relax
\mciteBstWouldAddEndPuncttrue
\mciteSetBstMidEndSepPunct{\mcitedefaultmidpunct}
{\mcitedefaultendpunct}{\mcitedefaultseppunct}\relax
\EndOfBibitem
\bibitem[van Aggelen \latin{et~al.}(2013)van Aggelen, Yang, and
  Yang]{van2013exchange}
van Aggelen,~H.; Yang,~Y.; Yang,~W. Exchange-correlation energy from pairing
  matrix fluctuation and the particle-particle random-phase approximation.
  \emph{Physical Review A} \textbf{2013}, \emph{88}, 030501\relax
\mciteBstWouldAddEndPuncttrue
\mciteSetBstMidEndSepPunct{\mcitedefaultmidpunct}
{\mcitedefaultendpunct}{\mcitedefaultseppunct}\relax
\EndOfBibitem
\bibitem[Yang \latin{et~al.}(2013)Yang, van Aggelen, and Yang]{yang2013double}
Yang,~Y.; van Aggelen,~H.; Yang,~W. Double, Rydberg and charge transfer
  excitations from pairing matrix fluctuation and particle-particle random
  phase approximation. \emph{The Journal of Chemical Physics} \textbf{2013},
  \emph{139}, 224105\relax
\mciteBstWouldAddEndPuncttrue
\mciteSetBstMidEndSepPunct{\mcitedefaultmidpunct}
{\mcitedefaultendpunct}{\mcitedefaultseppunct}\relax
\EndOfBibitem
\bibitem[Yang \latin{et~al.}(2015)Yang, Peng, Davidson, and Yang]{Yang2015}
Yang,~Y.; Peng,~D.; Davidson,~E.~R.; Yang,~W. {Singlet-triplet energy gaps for
  diradicals from particle-particle random phase approximation}. \emph{Journal
  of Physical Chemistry A} \textbf{2015}, \emph{119}, 4923--4932\relax
\mciteBstWouldAddEndPuncttrue
\mciteSetBstMidEndSepPunct{\mcitedefaultmidpunct}
{\mcitedefaultendpunct}{\mcitedefaultseppunct}\relax
\EndOfBibitem
\bibitem[Zhang \latin{et~al.}(2017)Zhang, Su, and Yang]{Zhang2017}
Zhang,~D.; Su,~N.~Q.; Yang,~W. {Accurate Quasiparticle Spectra from the
  T-Matrix Self-Energy and the Particle-Particle Random Phase Approximation}.
  \emph{Journal of Physical Chemistry Letters} \textbf{2017}, \emph{8},
  3223--3227\relax
\mciteBstWouldAddEndPuncttrue
\mciteSetBstMidEndSepPunct{\mcitedefaultmidpunct}
{\mcitedefaultendpunct}{\mcitedefaultseppunct}\relax
\EndOfBibitem
\bibitem[Scuseria \latin{et~al.}(2013)Scuseria, Henderson, and
  Bulik]{Scuseria2013}
Scuseria,~G.~E.; Henderson,~T.~M.; Bulik,~I.~W. {Particle-particle and
  quasiparticle random phase approximations: Connections to coupled cluster
  theory}. \emph{Journal of Chemical Physics} \textbf{2013}, \emph{139},
  104113\relax
\mciteBstWouldAddEndPuncttrue
\mciteSetBstMidEndSepPunct{\mcitedefaultmidpunct}
{\mcitedefaultendpunct}{\mcitedefaultseppunct}\relax
\EndOfBibitem
\bibitem[Peng \latin{et~al.}(2013)Peng, Steinmann, van Aggelen, and
  Yang]{peng2013equivalence}
Peng,~D.; Steinmann,~S.~N.; van Aggelen,~H.; Yang,~W. Equivalence of
  particle-particle random phase approximation correlation energy and
  ladder-coupled-cluster doubles. \emph{The Journal of Chemical Physics}
  \textbf{2013}, \emph{139}, 104112\relax
\mciteBstWouldAddEndPuncttrue
\mciteSetBstMidEndSepPunct{\mcitedefaultmidpunct}
{\mcitedefaultendpunct}{\mcitedefaultseppunct}\relax
\EndOfBibitem
\bibitem[Hohenstein \latin{et~al.}(2012)Hohenstein, Parrish, and
  Mart{\'{i}}nez]{Hohenstein2012}
Hohenstein,~E.~G.; Parrish,~R.~M.; Mart{\'{i}}nez,~T.~J. {Tensor
  hypercontraction density fitting. I. Quartic scaling second- and third-order
  M{\o}ller-Plesset perturbation theory}. \emph{Journal of Chemical Physics}
  \textbf{2012}, \emph{137}, 044103\relax
\mciteBstWouldAddEndPuncttrue
\mciteSetBstMidEndSepPunct{\mcitedefaultmidpunct}
{\mcitedefaultendpunct}{\mcitedefaultseppunct}\relax
\EndOfBibitem
\bibitem[Parrish \latin{et~al.}(2012)Parrish, Hohenstein, Mart{\'{i}}nez, and
  Sherrill]{Parrish2012}
Parrish,~R.~M.; Hohenstein,~E.~G.; Mart{\'{i}}nez,~T.~J.; Sherrill,~C.~D.
  {Tensor hypercontraction. II. Least-squares renormalization}. \emph{Journal
  of Chemical Physics} \textbf{2012}, \emph{137}, 224106\relax
\mciteBstWouldAddEndPuncttrue
\mciteSetBstMidEndSepPunct{\mcitedefaultmidpunct}
{\mcitedefaultendpunct}{\mcitedefaultseppunct}\relax
\EndOfBibitem
\bibitem[Hohenstein \latin{et~al.}(2012)Hohenstein, Parrish, Sherrill, and
  Mart{\'{i}}nez]{Hohenstein2012a}
Hohenstein,~E.~G.; Parrish,~R.~M.; Sherrill,~C.~D.; Mart{\'{i}}nez,~T.~J.
  {Communication: Tensor hypercontraction. III. Least-squares tensor
  hypercontraction for the determination of correlated wavefunctions}.
  \emph{Journal of Chemical Physics} \textbf{2012}, \emph{137}, 221101\relax
\mciteBstWouldAddEndPuncttrue
\mciteSetBstMidEndSepPunct{\mcitedefaultmidpunct}
{\mcitedefaultendpunct}{\mcitedefaultseppunct}\relax
\EndOfBibitem
\bibitem[Yeh and Morales(2023)Yeh, and Morales]{Yeh2023}
Yeh,~C.~N.; Morales,~M.~A. {Low-Scaling Algorithm for the Random Phase
  Approximation Using Tensor Hypercontraction with k-point Sampling}.
  \emph{Journal of Chemical Theory and Computation} \textbf{2023}, \emph{19},
  6197--6207\relax
\mciteBstWouldAddEndPuncttrue
\mciteSetBstMidEndSepPunct{\mcitedefaultmidpunct}
{\mcitedefaultendpunct}{\mcitedefaultseppunct}\relax
\EndOfBibitem
\bibitem[Duchemin and Blase(2019)Duchemin, and Blase]{Duchemin2019}
Duchemin,~I.; Blase,~X. {Separable resolution-of-the-identity with all-electron
  Gaussian bases: Application to cubic-scaling RPA}. \emph{Journal of Chemical
  Physics} \textbf{2019}, \emph{150}, 174120\relax
\mciteBstWouldAddEndPuncttrue
\mciteSetBstMidEndSepPunct{\mcitedefaultmidpunct}
{\mcitedefaultendpunct}{\mcitedefaultseppunct}\relax
\EndOfBibitem
\bibitem[Duchemin and Blase(2021)Duchemin, and Blase]{Duchemin2021a}
Duchemin,~I.; Blase,~X. {Cubic-Scaling All-Electron GW Calculations with a
  Separable Density-Fitting Space-Time Approach}. \emph{Journal of Chemical
  Theory and Computation} \textbf{2021}, \emph{17}, 2383--2393\relax
\mciteBstWouldAddEndPuncttrue
\mciteSetBstMidEndSepPunct{\mcitedefaultmidpunct}
{\mcitedefaultendpunct}{\mcitedefaultseppunct}\relax
\EndOfBibitem
\bibitem[Yeh and Morales(2024)Yeh, and Morales]{Yeh2024a}
Yeh,~C.~N.; Morales,~M.~A. {Low-Scaling Algorithms for GW and Constrained
  Random Phase Approximation Using Symmetry-Adapted Interpolative Separable
  Density Fitting}. \emph{Journal of Chemical Theory and Computation}
  \textbf{2024}, \emph{20}, 3184--3198\relax
\mciteBstWouldAddEndPuncttrue
\mciteSetBstMidEndSepPunct{\mcitedefaultmidpunct}
{\mcitedefaultendpunct}{\mcitedefaultseppunct}\relax
\EndOfBibitem
\bibitem[Shenvi \latin{et~al.}(2014)Shenvi, van Aggelen, Yang, and
  Yang]{Shenvi2014Tensorr4}
Shenvi,~N.; van Aggelen,~H.; Yang,~Y.; Yang,~W. {Tensor hypercontracted ppRPA:
  Reducing the cost of the particle-particle random phase approximation from O
  (r6) to O (r4)}. \emph{Journal of Chemical Physics} \textbf{2014},
  \emph{141}, 024119\relax
\mciteBstWouldAddEndPuncttrue
\mciteSetBstMidEndSepPunct{\mcitedefaultmidpunct}
{\mcitedefaultendpunct}{\mcitedefaultseppunct}\relax
\EndOfBibitem
\bibitem[Meyer(1973)]{meyer1973pno}
Meyer,~W. PNO--CI Studies of electron correlation effects. I. Configuration
  expansion by means of nonorthogonal orbitals, and application to the ground
  state and ionized states of methane. \emph{The Journal of Chemical Physics}
  \textbf{1973}, \emph{58}, 1017--1035\relax
\mciteBstWouldAddEndPuncttrue
\mciteSetBstMidEndSepPunct{\mcitedefaultmidpunct}
{\mcitedefaultendpunct}{\mcitedefaultseppunct}\relax
\EndOfBibitem
\bibitem[Ahlrichs(1979)]{ahlrichs1979many}
Ahlrichs,~R. Many body perturbation calculations and coupled electron pair
  models. \emph{Computer Physics Communications} \textbf{1979}, \emph{17},
  31--45\relax
\mciteBstWouldAddEndPuncttrue
\mciteSetBstMidEndSepPunct{\mcitedefaultmidpunct}
{\mcitedefaultendpunct}{\mcitedefaultseppunct}\relax
\EndOfBibitem
\bibitem[Carter-Fenk(2025)]{Carter-Fenk2025DiagrammaticTheory}
Carter-Fenk,~K. {Diagrammatic Simplification of Linearized Coupled Cluster
  Theory}. \emph{Journal of Physical Chemistry A} \textbf{2025}, \emph{129},
  7251--7260\relax
\mciteBstWouldAddEndPuncttrue
\mciteSetBstMidEndSepPunct{\mcitedefaultmidpunct}
{\mcitedefaultendpunct}{\mcitedefaultseppunct}\relax
\EndOfBibitem
\bibitem[Faddeev(1961)]{faddeev2016scattering}
Faddeev,~L.~D. Scattering theory for a three-particle system. \emph{Soviet
  Physics JETP} \textbf{1961}, \emph{12}, 1014--1019, Russian original: Zh.
  Eksp. Teor. Fiz. \textbf{39}, 1459 (1960). Reprinted in \emph{Fifty Years of
  Mathematical Physics: Selected Works of Ludwig Faddeev}; World Scientific,
  2016; pp 37--42\relax
\mciteBstWouldAddEndPuncttrue
\mciteSetBstMidEndSepPunct{\mcitedefaultmidpunct}
{\mcitedefaultendpunct}{\mcitedefaultseppunct}\relax
\EndOfBibitem
\bibitem[Degroote \latin{et~al.}(2011)Degroote, Van~Neck, and
  Barbieri]{Degroote2011}
Degroote,~M.; Van~Neck,~D.; Barbieri,~C. {Faddeev random-phase approximation
  for molecules}. \emph{Physical Review A} \textbf{2011}, \emph{83},
  042517\relax
\mciteBstWouldAddEndPuncttrue
\mciteSetBstMidEndSepPunct{\mcitedefaultmidpunct}
{\mcitedefaultendpunct}{\mcitedefaultseppunct}\relax
\EndOfBibitem
\bibitem[Coveney(2025)]{Coveney2025UncoveringTheory}
Coveney,~C.~J. {Uncovering Relationships between the Electronic Self-Energy and
  Coupled-Cluster Doubles Theory}. \emph{Journal of Physical Chemistry A}
  \textbf{2025}, \emph{129}, 8689--8698\relax
\mciteBstWouldAddEndPuncttrue
\mciteSetBstMidEndSepPunct{\mcitedefaultmidpunct}
{\mcitedefaultendpunct}{\mcitedefaultseppunct}\relax
\EndOfBibitem
\bibitem[Hodecker \latin{et~al.}(2019)Hodecker, Dempwolff, Rehn, and
  Dreuw]{Hodecker2019Algebraic-diagrammaticEnergies}
Hodecker,~M.; Dempwolff,~A.~L.; Rehn,~D.~R.; Dreuw,~A. {Algebraic-diagrammatic
  construction scheme for the polarization propagator including ground-state
  coupled-cluster amplitudes. I. Excitation energies}. \emph{Journal of
  Chemical Physics} \textbf{2019}, \emph{150}, 174104\relax
\mciteBstWouldAddEndPuncttrue
\mciteSetBstMidEndSepPunct{\mcitedefaultmidpunct}
{\mcitedefaultendpunct}{\mcitedefaultseppunct}\relax
\EndOfBibitem
\bibitem[Hodecker \latin{et~al.}(2019)Hodecker, Rehn, Norman, and
  Dreuw]{Hodecker2019Algebraic-diagrammaticPolarizabilities}
Hodecker,~M.; Rehn,~D.~R.; Norman,~P.; Dreuw,~A. {Algebraic-diagrammatic
  construction scheme for the polarization propagator including ground-state
  coupled-cluster amplitudes. II. Static polarizabilities}. \emph{Journal of
  Chemical Physics} \textbf{2019}, \emph{150}, 174105\relax
\mciteBstWouldAddEndPuncttrue
\mciteSetBstMidEndSepPunct{\mcitedefaultmidpunct}
{\mcitedefaultendpunct}{\mcitedefaultseppunct}\relax
\EndOfBibitem
\bibitem[Schirmer \latin{et~al.}(1998)Schirmer, Trofimov, and
  Stelter]{Schirmer1998APropagator}
Schirmer,~J.; Trofimov,~A.~B.; Stelter,~G. {A non-Dyson third-order
  approximation scheme for the electron propagator}. \emph{Journal of Chemical
  Physics} \textbf{1998}, \emph{109}, 4734--4744\relax
\mciteBstWouldAddEndPuncttrue
\mciteSetBstMidEndSepPunct{\mcitedefaultmidpunct}
{\mcitedefaultendpunct}{\mcitedefaultseppunct}\relax
\EndOfBibitem
\bibitem[Dempwolff \latin{et~al.}(2019)Dempwolff, Schneider, Hodecker, and
  Dreuw]{Dempwolff2019EfficientMolecules}
Dempwolff,~A.~L.; Schneider,~M.; Hodecker,~M.; Dreuw,~A. {Efficient
  implementation of the non-Dyson third-order algebraic diagrammatic
  construction approximation for the electron propagator for closed- and
  open-shell molecules}. \emph{Journal of Chemical Physics} \textbf{2019},
  \emph{150}, 064108\relax
\mciteBstWouldAddEndPuncttrue
\mciteSetBstMidEndSepPunct{\mcitedefaultmidpunct}
{\mcitedefaultendpunct}{\mcitedefaultseppunct}\relax
\EndOfBibitem
\bibitem[Banerjee and Sokolov(2019)Banerjee, and
  Sokolov]{Banerjee2019Third-orderImplementation}
Banerjee,~S.; Sokolov,~A.~Y. {Third-order algebraic diagrammatic construction
  theory for electron attachment and ionization energies: Conventional and
  Green's function implementation}. \emph{Journal of Chemical Physics}
  \textbf{2019}, \emph{151}, 224112\relax
\mciteBstWouldAddEndPuncttrue
\mciteSetBstMidEndSepPunct{\mcitedefaultmidpunct}
{\mcitedefaultendpunct}{\mcitedefaultseppunct}\relax
\EndOfBibitem
\bibitem[Herbst \latin{et~al.}(2020)Herbst, Scheurer, Fransson, Rehn, and
  Dreuw]{Herbst2020Adcc:Methods}
Herbst,~M.~F.; Scheurer,~M.; Fransson,~T.; Rehn,~D.~R.; Dreuw,~A. {adcc: A
  versatile toolkit for rapid development of algebraic-diagrammatic
  construction methods}. \emph{Wiley Interdisciplinary Reviews: Computational
  Molecular Science} \textbf{2020}, \emph{10}, e1462\relax
\mciteBstWouldAddEndPuncttrue
\mciteSetBstMidEndSepPunct{\mcitedefaultmidpunct}
{\mcitedefaultendpunct}{\mcitedefaultseppunct}\relax
\EndOfBibitem
\bibitem[Trofimov and Schirmer(2005)Trofimov, and
  Schirmer]{Trofimov2005MolecularApproach}
Trofimov,~A.~B.; Schirmer,~J. {Molecular ionization energies and ground- and
  ionic-state properties using a non-Dyson electron propagator approach}.
  \emph{Journal of Chemical Physics} \textbf{2005}, \emph{123}, 144115\relax
\mciteBstWouldAddEndPuncttrue
\mciteSetBstMidEndSepPunct{\mcitedefaultmidpunct}
{\mcitedefaultendpunct}{\mcitedefaultseppunct}\relax
\EndOfBibitem
\bibitem[Keizer \latin{et~al.}(2026)Keizer, Paggi, Berger, Romaniello, and
  F{\"{o}}rster]{Keizer2026BenchmarkMolecules}
Keizer,~M.; Paggi,~S.; Berger,~J.~A.; Romaniello,~P.; F{\"{o}}rster,~A.
  {Benchmark of Multi-Channel Dyson Equation and Algebraic Diagrammatic
  Construction Methods for molecules}. \emph{arXiv:2608.25669} \textbf{2026},
  1--25\relax
\mciteBstWouldAddEndPuncttrue
\mciteSetBstMidEndSepPunct{\mcitedefaultmidpunct}
{\mcitedefaultendpunct}{\mcitedefaultseppunct}\relax
\EndOfBibitem
\bibitem[Bruneval \latin{et~al.}(2026)Bruneval, Verzijl, F{\"{o}}rster, and
  Rodriguez-Mayorga]{Bruneval2026GWEquation}
Bruneval,~F.; Verzijl,~E.; F{\"{o}}rster,~A.; Rodriguez-Mayorga,~M. {GW reduced
  density matrix from iterated linearized Dyson equation}.
  \emph{arXiv:2607.15695} \textbf{2026}, 1--9\relax
\mciteBstWouldAddEndPuncttrue
\mciteSetBstMidEndSepPunct{\mcitedefaultmidpunct}
{\mcitedefaultendpunct}{\mcitedefaultseppunct}\relax
\EndOfBibitem
\bibitem[Sun \latin{et~al.}(2020)Sun, Zhang, Banerjee, Bao, Barbry, Blunt,
  Bogdanov, Booth, Chen, Cui, Eriksen, Gao, Guo, Hermann, Hermes, Koh, Koval,
  Lehtola, Li, Liu, Mardirossian, McClain, Motta, Mussard, Pham, Pulkin,
  Purwanto, Robinson, Ronca, Sayfutyarova, Scheurer, Schurkus, Smith, Sun, Sun,
  Upadhyay, Wagner, Wang, White, Whitfield, Williamson, Wouters, Yang, Yu, Zhu,
  Berkelbach, Sharma, Sokolov, and Chan]{Sun2020}
Sun,~Q.; Zhang,~X.; Banerjee,~S.; Bao,~P.; Barbry,~M.; Blunt,~N.~S.;
  Bogdanov,~N.~A.; Booth,~G.~H.; Chen,~J.; Cui,~Z.~H.; Eriksen,~J.~J.; Gao,~Y.;
  Guo,~S.; Hermann,~J.; Hermes,~M.~R.; Koh,~K.; Koval,~P.; Lehtola,~S.; Li,~Z.;
  Liu,~J.; Mardirossian,~N.; McClain,~J.~D.; Motta,~M.; Mussard,~B.;
  Pham,~H.~Q.; Pulkin,~A.; Purwanto,~W.; Robinson,~P.~J.; Ronca,~E.;
  Sayfutyarova,~E.~R.; Scheurer,~M.; Schurkus,~H.~F.; Smith,~J.~E.; Sun,~C.;
  Sun,~S.~N.; Upadhyay,~S.; Wagner,~L.~K.; Wang,~X.; White,~A.;
  Whitfield,~J.~D.; Williamson,~M.~J.; Wouters,~S.; Yang,~J.; Yu,~J.~M.;
  Zhu,~T.; Berkelbach,~T.~C.; Sharma,~S.; Sokolov,~A.~Y.; Chan,~G. K.~L.
  {Recent developments in the PySCF program package}. \emph{Journal of Chemical
  Physics} \textbf{2020}, \emph{153}, 024109\relax
\mciteBstWouldAddEndPuncttrue
\mciteSetBstMidEndSepPunct{\mcitedefaultmidpunct}
{\mcitedefaultendpunct}{\mcitedefaultseppunct}\relax
\EndOfBibitem
\bibitem[Sun \latin{et~al.}(2026)Sun, Hermes, Wu, Zhai, Zhang, Ahmed, Aucar,
  Backhouse, Banerjee, Bao, Bogdanov, Bystrom, Chapoton, Chen, Chernyshov,
  Clifford, Cohen-Janes, Cui, Damour, Dattani, Dittmer, Ehlert, Eriksen,
  Evangelista, Ewing, Farahvash, Focke, Gao, Gasperich, Gillispie, Greiner,
  Hennefarth, Hermann, Hillenbrand, Huhtasalo, Ibrahim, Jangid, Javaremi,
  Jenkins, Jin, King, Kooi, Kurian, Larsson, Lau, Lee, Lehtola, Li, Li, Li, Li,
  Li, Lykhin, Mahajan, Mauger, del Mazo-Sevillano, Moussa, Nakano, Neufeld,
  Peng, Pham, Pinski, Pokhilko, Pu, Qian, Quiton, Schulze, Scott, Seal, Serna,
  Smith, Smyser, Stahl, Sun, Sung, Trushin, Upadhyay, Vo, Vogels, Wang, Wang,
  Wang, Wang, Wang, Williamson, Yang, Ye, Yeh, Yu, Yu, Yu, Zhang, Zhang, Zhang,
  Zhao, Zhou, Zhu, Zhu, Berkelbach, Gagliardi, Sharma, Sokolov, and
  Chan]{Sun2026TheProject}
Sun,~Q.; Hermes,~M.~R.; Wu,~X.; Zhai,~H.; Zhang,~X.; Ahmed,~A.~M.;
  Aucar,~J.~J.; Backhouse,~O.~J.; Banerjee,~S.; Bao,~P.; Bogdanov,~N.~A.;
  Bystrom,~K.; Chapoton,~F.; Chen,~N.-Y.; Chernyshov,~I.~Y.; Clifford,~H.~S.;
  Cohen-Janes,~S.; Cui,~Z.-H.; Damour,~Y.~D.; Dattani,~N.; Dittmer,~L.~B.;
  Ehlert,~S.; Eriksen,~J.~J.; Evangelista,~F.~A.; Ewing,~S.~A.; Farahvash,~A.;
  Focke,~K.; Gao,~Y.; Gasperich,~K.~E.; Gillispie,~N.; Greiner,~J.;
  Hennefarth,~M.~R.; Hermann,~J.; Hillenbrand,~C.; Huhtasalo,~J.; Ibrahim,~B.;
  Jangid,~B.; Javaremi,~A.~N.; Jenkins,~A.~J.; Jin,~Y.; King,~D.~S.;
  Kooi,~D.~P.; Kurian,~J.~S.; Larsson,~H.~R.; Lau,~B. T.~G.; Lee,~S.;
  Lehtola,~S.; Li,~C.; Li,~H.; Li,~J.; Li,~R.; Li,~S.; Lykhin,~A.~O.;
  Mahajan,~A.; Mauger,~N.; del Mazo-Sevillano,~P.; Moussa,~J.; Nakano,~K.;
  Neufeld,~V.~A.; Peng,~L.; Pham,~H.~Q.; Pinski,~P.; Pokhilko,~P.; Pu,~Z.;
  Qian,~Y.; Quiton,~S.~J.; Schulze,~W.~T.; Scott,~T.~R.; Seal,~A.;
  Serna,~J.~D.; Smith,~J. E.~T.; Smyser,~K.~E.; Stahl,~T.; Sun,~C.;
  Sung,~K.~J.; Trushin,~E.; Upadhyay,~S.; Vo,~E.~A.; Vogels,~T.; Wang,~S.;
  Wang,~T.; Wang,~X.; Wang,~X.; Wang,~Y.; Williamson,~M.; Yang,~J.; Ye,~H.-Z.;
  Yeh,~C.-N.; Yu,~H.; Yu,~J.; Yu,~V. W.-z.; Zhang,~C.; Zhang,~D.; Zhang,~Y.;
  Zhao,~Z.; Zhou,~Z.; Zhu,~A.~J.; Zhu,~T.; Berkelbach,~T.~C.; Gagliardi,~L.;
  Sharma,~S.; Sokolov,~A.; Chan,~G. K.-L. {The Python Simulations of Chemistry
  Framework: 10 years of an open-source quantum chemistry project}.
  \emph{arXiv:2603.14155} \textbf{2026}, \relax
\mciteBstWouldAddEndPunctfalse
\mciteSetBstMidEndSepPunct{\mcitedefaultmidpunct}
{}{\mcitedefaultseppunct}\relax
\EndOfBibitem
\bibitem[Rubin and DePrince~III(2021)Rubin, and DePrince~III]{rubin2021p}
Rubin,~N.~C.; DePrince~III,~A.~E. p† q: a tool for prototyping many-body
  methods for quantum chemistry. \emph{Molecular Physics} \textbf{2021},
  \emph{119}, e1954709\relax
\mciteBstWouldAddEndPuncttrue
\mciteSetBstMidEndSepPunct{\mcitedefaultmidpunct}
{\mcitedefaultendpunct}{\mcitedefaultseppunct}\relax
\EndOfBibitem
\bibitem[Liebenthal \latin{et~al.}(2025)Liebenthal, Yuwono, Koulias, Li, Rubin,
  and DePrince~III]{Liebenthal2025AutomatedJCTC}
Liebenthal,~M.~D.; Yuwono,~S.~H.; Koulias,~L.~N.; Li,~R.~R.; Rubin,~N.~C.;
  DePrince~III,~A.~E. Automated quantum chemistry code generation with the p†
  q package. \emph{The Journal of Physical Chemistry A} \textbf{2025},
  \emph{129}, 6679--6693\relax
\mciteBstWouldAddEndPuncttrue
\mciteSetBstMidEndSepPunct{\mcitedefaultmidpunct}
{\mcitedefaultendpunct}{\mcitedefaultseppunct}\relax
\EndOfBibitem
\bibitem[Stahl \latin{et~al.}(2022)Stahl, Banerjee, and
  Sokolov]{Stahl2022QuantifyingExcitations}
Stahl,~T.~L.; Banerjee,~S.; Sokolov,~A.~Y. {Quantifying and reducing spin
  contamination in algebraic diagrammatic construction theory of charged
  excitations}. \emph{Journal of Chemical Physics} \textbf{2022}, \emph{157},
  044106\relax
\mciteBstWouldAddEndPuncttrue
\mciteSetBstMidEndSepPunct{\mcitedefaultmidpunct}
{\mcitedefaultendpunct}{\mcitedefaultseppunct}\relax
\EndOfBibitem
\bibitem[Dunning(1989)]{Dunning1989}
Dunning,~T.~H. {Gaussian basis sets for use in correlated molecular
  calculations. I. The atoms boron through neon and hydrogen}. \emph{The
  Journal of Chemical Physics} \textbf{1989}, \emph{90}, 1007--1023\relax
\mciteBstWouldAddEndPuncttrue
\mciteSetBstMidEndSepPunct{\mcitedefaultmidpunct}
{\mcitedefaultendpunct}{\mcitedefaultseppunct}\relax
\EndOfBibitem
\bibitem[Kendall \latin{et~al.}(1992)Kendall, Dunning, and
  Harrison]{Kendall1992}
Kendall,~R.~A.; Dunning,~T.~H.; Harrison,~R.~J. {Electron affinities of the
  first-row atoms revisited. Systematic basis sets and wave functions}.
  \emph{The Journal of Chemical Physics} \textbf{1992}, \emph{96},
  6796--6806\relax
\mciteBstWouldAddEndPuncttrue
\mciteSetBstMidEndSepPunct{\mcitedefaultmidpunct}
{\mcitedefaultendpunct}{\mcitedefaultseppunct}\relax
\EndOfBibitem
\bibitem[Woon and Dunning(1993)Woon, and Dunning]{Woon1993}
Woon,~D.~E.; Dunning,~T.~H. {Gaussian basis sets for use in correlated
  molecular calculations. III. The atoms aluminum through argon}. \emph{The
  Journal of Chemical Physics} \textbf{1993}, \emph{98}, 1358--1371\relax
\mciteBstWouldAddEndPuncttrue
\mciteSetBstMidEndSepPunct{\mcitedefaultmidpunct}
{\mcitedefaultendpunct}{\mcitedefaultseppunct}\relax
\EndOfBibitem
\bibitem[Weigend \latin{et~al.}(2002)Weigend, K{\"{o}}hn, and
  H{\"{a}}ttig]{Weigend2002}
Weigend,~F.; K{\"{o}}hn,~A.; H{\"{a}}ttig,~C. {Efficient use of the correlation
  consistent basis sets in resolution of the identity MP2 calculations}.
  \emph{Journal of Chemical Physics} \textbf{2002}, \emph{116},
  3175--3183\relax
\mciteBstWouldAddEndPuncttrue
\mciteSetBstMidEndSepPunct{\mcitedefaultmidpunct}
{\mcitedefaultendpunct}{\mcitedefaultseppunct}\relax
\EndOfBibitem
\bibitem[H{\"{a}}ttig(2005)]{Hattig2005}
H{\"{a}}ttig,~C. {Optimization of auxiliary basis sets for RI-MP2 and RI-CC2
  calculations: Core-valence and quintuple-{$\zeta$} basis sets for H to Ar and
  QZVPP basis sets for Li to Kr}. \emph{Physical Chemistry Chemical Physics}
  \textbf{2005}, \emph{7}, 59--66\relax
\mciteBstWouldAddEndPuncttrue
\mciteSetBstMidEndSepPunct{\mcitedefaultmidpunct}
{\mcitedefaultendpunct}{\mcitedefaultseppunct}\relax
\EndOfBibitem
\bibitem[Stephens \latin{et~al.}(1994)Stephens, Devlin, Chabalowski, and
  Frisch]{Stephens1994}
Stephens,~P.~J.; Devlin,~F.~J.; Chabalowski,~C.~F.; Frisch,~M.~J. {Ab Initio
  calculation of vibrational absorption and circular dichroism spectra using
  density functional force fields}. \emph{Journal of Physical Chemistry}
  \textbf{1994}, \emph{98}, 11623--11627\relax
\mciteBstWouldAddEndPuncttrue
\mciteSetBstMidEndSepPunct{\mcitedefaultmidpunct}
{\mcitedefaultendpunct}{\mcitedefaultseppunct}\relax
\EndOfBibitem
\bibitem[Weigend and Ahlrichs(2005)Weigend, and Ahlrichs]{Weigend2005}
Weigend,~F.; Ahlrichs,~R. {Balanced basis sets of split valence, triple zeta
  valence and quadruple zeta valence quality for H to Rn: Design and assessment
  of accuracy}. \emph{Physical Chemistry Chemical Physics} \textbf{2005},
  \emph{7}, 3297--3305\relax
\mciteBstWouldAddEndPuncttrue
\mciteSetBstMidEndSepPunct{\mcitedefaultmidpunct}
{\mcitedefaultendpunct}{\mcitedefaultseppunct}\relax
\EndOfBibitem
\bibitem[Marie and Loos(2024)Marie, and Loos]{Marie2024}
Marie,~A.; Loos,~P.~F. {Reference Energies for Valence Ionizations and
  Satellite Transitions}. \emph{Journal of Chemical Theory and Computation}
  \textbf{2024}, \emph{20}, 4751–4777\relax
\mciteBstWouldAddEndPuncttrue
\mciteSetBstMidEndSepPunct{\mcitedefaultmidpunct}
{\mcitedefaultendpunct}{\mcitedefaultseppunct}\relax
\EndOfBibitem
\bibitem[Huron \latin{et~al.}(1973)Huron, Malrieu, and Rancurel]{Huron1973}
Huron,~B.; Malrieu,~J.~P.; Rancurel,~P. {Iterative perturbation calculations of
  ground and excited state energies from multiconfigurational zeroth-order wave
  functions}. \emph{The Journal of Chemical Physics} \textbf{1973}, \emph{58},
  5745--5759\relax
\mciteBstWouldAddEndPuncttrue
\mciteSetBstMidEndSepPunct{\mcitedefaultmidpunct}
{\mcitedefaultendpunct}{\mcitedefaultseppunct}\relax
\EndOfBibitem
\bibitem[Garniron \latin{et~al.}(2019)Garniron, Applencourt, Gasperich, Benali,
  Fert{\'{e}}, Paquier, Pradines, Assaraf, Reinhardt, Toulouse, Barbaresco,
  Renon, David, Malrieu, V{\'{e}}ril, Caffarel, Loos, Giner, and
  Scemama]{Garniron2019QuantumPrograms}
Garniron,~Y.; Applencourt,~T.; Gasperich,~K.; Benali,~A.; Fert{\'{e}},~A.;
  Paquier,~J.; Pradines,~B.; Assaraf,~R.; Reinhardt,~P.; Toulouse,~J.;
  Barbaresco,~P.; Renon,~N.; David,~G.; Malrieu,~J.~P.; V{\'{e}}ril,~M.;
  Caffarel,~M.; Loos,~P.~F.; Giner,~E.; Scemama,~A. {Quantum Package 2.0: An
  Open-Source Determinant-Driven Suite of Programs}. \emph{Journal of Chemical
  Theory and Computation} \textbf{2019}, \emph{15}, 3591--3609\relax
\mciteBstWouldAddEndPuncttrue
\mciteSetBstMidEndSepPunct{\mcitedefaultmidpunct}
{\mcitedefaultendpunct}{\mcitedefaultseppunct}\relax
\EndOfBibitem
\bibitem[Strinati(1988)]{Strinati1988}
Strinati,~G. {Application of the Green's functions method to the study of the
  optical properties of semiconductors}. \emph{La Rivista Del Nuovo Cimento
  Series 3} \textbf{1988}, \emph{11}, 1--86\relax
\mciteBstWouldAddEndPuncttrue
\mciteSetBstMidEndSepPunct{\mcitedefaultmidpunct}
{\mcitedefaultendpunct}{\mcitedefaultseppunct}\relax
\EndOfBibitem
\bibitem[Rohlfing and Louie(2000)Rohlfing, and Louie]{Rohlfing2000}
Rohlfing,~M.; Louie,~S.~G. {Electron-hole excitations and optical spectra from
  first principles}. \emph{Physical Review B} \textbf{2000}, \emph{62},
  4927--4944\relax
\mciteBstWouldAddEndPuncttrue
\mciteSetBstMidEndSepPunct{\mcitedefaultmidpunct}
{\mcitedefaultendpunct}{\mcitedefaultseppunct}\relax
\EndOfBibitem
\bibitem[Onida \latin{et~al.}(2002)Onida, Reining, and Rubio]{Onida2002}
Onida,~G.; Reining,~L.; Rubio,~A. {Electronic excitations: density-functional
  versus many-body Green’s-function approaches}. \emph{Reviews of Modern
  Physics} \textbf{2002}, \emph{74}, 601\relax
\mciteBstWouldAddEndPuncttrue
\mciteSetBstMidEndSepPunct{\mcitedefaultmidpunct}
{\mcitedefaultendpunct}{\mcitedefaultseppunct}\relax
\EndOfBibitem
\bibitem[Blase \latin{et~al.}(2018)Blase, Duchemin, and Jacquemin]{Blase2018}
Blase,~X.; Duchemin,~I.; Jacquemin,~D. {The Bethe-Salpeter equation in
  chemistry: Relations with TD-DFT, applications and challenges}.
  \emph{Chemical Society Reviews} \textbf{2018}, \emph{47}, 1022--1043\relax
\mciteBstWouldAddEndPuncttrue
\mciteSetBstMidEndSepPunct{\mcitedefaultmidpunct}
{\mcitedefaultendpunct}{\mcitedefaultseppunct}\relax
\EndOfBibitem
\bibitem[Cederbaum \latin{et~al.}(1977)Cederbaum, Schirmer, Domcke, and
  Von~Niessen]{Cederbaum1977}
Cederbaum,~L.~S.; Schirmer,~J.; Domcke,~W.; Von~Niessen,~W. {Complete breakdown
  of the quasiparticle picture for inner valence electrons}. \emph{Journal of
  Physics B: Atomic and Molecular Physics} \textbf{1977}, \emph{10},
  L549--L553\relax
\mciteBstWouldAddEndPuncttrue
\mciteSetBstMidEndSepPunct{\mcitedefaultmidpunct}
{\mcitedefaultendpunct}{\mcitedefaultseppunct}\relax
\EndOfBibitem
\bibitem[Mejuto-Zaera \latin{et~al.}(2021)Mejuto-Zaera, Weng, Romanova, Cotton,
  Whaley, Tubman, and Vl{\v{c}}ek]{Mejuto-Zaera2021}
Mejuto-Zaera,~C.; Weng,~G.; Romanova,~M.; Cotton,~S.~J.; Whaley,~K.~B.;
  Tubman,~N.~M.; Vl{\v{c}}ek,~V. {Are multi-quasiparticle interactions
  important in molecular ionization?} \emph{Journal of Chemical Physics}
  \textbf{2021}, \emph{154}, 121101\relax
\mciteBstWouldAddEndPuncttrue
\mciteSetBstMidEndSepPunct{\mcitedefaultmidpunct}
{\mcitedefaultendpunct}{\mcitedefaultseppunct}\relax
\EndOfBibitem
\bibitem[Pavlyukh \latin{et~al.}(2026)Pavlyukh, Bruneval, and
  F{\"{o}}rster]{Pavlyukh2026ApproachingSelf-Energies}
Pavlyukh,~Y.; Bruneval,~F.; F{\"{o}}rster,~A. {Approaching Coupled Cluster
  Accuracy with Positive Semidefinite Vertex Corrected Self-Energies}.
  \emph{arXiv:2607.29359} \textbf{2026}, 1--17\relax
\mciteBstWouldAddEndPuncttrue
\mciteSetBstMidEndSepPunct{\mcitedefaultmidpunct}
{\mcitedefaultendpunct}{\mcitedefaultseppunct}\relax
\EndOfBibitem
\bibitem[Edmiston and Ruedenberg(1965)Edmiston, and Ruedenberg]{Edmiston1965a}
Edmiston,~C.; Ruedenberg,~K. {Localized Atomic and Molecular Orbitals. II}.
  \emph{Journal of Chemical Physics} \textbf{1965}, \emph{43}, 97--116\relax
\mciteBstWouldAddEndPuncttrue
\mciteSetBstMidEndSepPunct{\mcitedefaultmidpunct}
{\mcitedefaultendpunct}{\mcitedefaultseppunct}\relax
\EndOfBibitem
\bibitem[Pickup and Goscinski(1973)Pickup, and
  Goscinski]{Pickup1973DirectShells}
Pickup,~B.~T.; Goscinski,~O. {Direct calculation of ionization energies I.
  Closed shells}. \emph{Molecular Physics} \textbf{1973}, \emph{26},
  1013--1035\relax
\mciteBstWouldAddEndPuncttrue
\mciteSetBstMidEndSepPunct{\mcitedefaultmidpunct}
{\mcitedefaultendpunct}{\mcitedefaultseppunct}\relax
\EndOfBibitem
\bibitem[Halkier \latin{et~al.}(1998)Halkier, Helgaker, J{\o}rgensen, Klopper,
  Koch, Olsen, and Wilson]{Halkier1998}
Halkier,~A.; Helgaker,~T.; J{\o}rgensen,~P.; Klopper,~W.; Koch,~H.; Olsen,~J.;
  Wilson,~A.~K. {Basis-set convergence in correlated calculations on Ne, N2 ,
  and H2O}. \emph{Chemical Physics Letters} \textbf{1998}, \emph{286},
  243--252\relax
\mciteBstWouldAddEndPuncttrue
\mciteSetBstMidEndSepPunct{\mcitedefaultmidpunct}
{\mcitedefaultendpunct}{\mcitedefaultseppunct}\relax
\EndOfBibitem
\bibitem[Hachmann \latin{et~al.}(2007)Hachmann, Dorando, Avil{\'{e}}s, and
  Chan]{Hachmann2007}
Hachmann,~J.; Dorando,~J.~J.; Avil{\'{e}}s,~M.; Chan,~G. K.~L. {The radical
  character of the acenes: A density matrix renormalization group study}.
  \emph{Journal of Chemical Physics} \textbf{2007}, \emph{127}, 134309\relax
\mciteBstWouldAddEndPuncttrue
\mciteSetBstMidEndSepPunct{\mcitedefaultmidpunct}
{\mcitedefaultendpunct}{\mcitedefaultseppunct}\relax
\EndOfBibitem
\bibitem[Hajgat{\'{o}} \latin{et~al.}(2009)Hajgat{\'{o}}, Szieberth, Geerlings,
  De~Proft, and Deleuze]{Hajgato2009}
Hajgat{\'{o}},~B.; Szieberth,~D.; Geerlings,~P.; De~Proft,~F.; Deleuze,~M.~S.
  {A benchmark theoretical study of the electronic ground state and of the
  singlet-triplet split of benzene and linear acenes}. \emph{Journal of
  Chemical Physics} \textbf{2009}, \emph{131}, 224321\relax
\mciteBstWouldAddEndPuncttrue
\mciteSetBstMidEndSepPunct{\mcitedefaultmidpunct}
{\mcitedefaultendpunct}{\mcitedefaultseppunct}\relax
\EndOfBibitem
\bibitem[Hajgat{\'{o}} \latin{et~al.}(2011)Hajgat{\'{o}}, Huzak, and
  Deleuze]{Hajgato2011}
Hajgat{\'{o}},~B.; Huzak,~M.; Deleuze,~M.~S. {Focal point analysis of the
  singlet-triplet energy gap of octacene and larger acenes}. \emph{Journal of
  Physical Chemistry A} \textbf{2011}, \emph{115}, 9282--9293\relax
\mciteBstWouldAddEndPuncttrue
\mciteSetBstMidEndSepPunct{\mcitedefaultmidpunct}
{\mcitedefaultendpunct}{\mcitedefaultseppunct}\relax
\EndOfBibitem
\bibitem[Zimmerman(2017)]{Zimmerman2017}
Zimmerman,~P.~M. {Singlet-Triplet Gaps through Incremental Full Configuration
  Interaction}. \emph{Journal of Physical Chemistry A} \textbf{2017},
  \emph{121}, 4712--4720\relax
\mciteBstWouldAddEndPuncttrue
\mciteSetBstMidEndSepPunct{\mcitedefaultmidpunct}
{\mcitedefaultendpunct}{\mcitedefaultseppunct}\relax
\EndOfBibitem
\bibitem[Ghosh \latin{et~al.}(2017)Ghosh, Cramer, Truhlar, and
  Gagliardi]{Ghosh2017}
Ghosh,~S.; Cramer,~C.~J.; Truhlar,~D.~G.; Gagliardi,~L.
  {Generalized-active-space pair-density functional theory: an efficient method
  to study large, strongly correlated, conjugated systems}. \emph{Chemical
  Science} \textbf{2017}, \emph{8}, 2741--2750\relax
\mciteBstWouldAddEndPuncttrue
\mciteSetBstMidEndSepPunct{\mcitedefaultmidpunct}
{\mcitedefaultendpunct}{\mcitedefaultseppunct}\relax
\EndOfBibitem
\bibitem[Mostafanejad and DePrince~III(2019)Mostafanejad, and
  DePrince~III]{Mostafanejad2019}
Mostafanejad,~M.; DePrince~III,~A.~E. {Combining Pair-Density Functional Theory
  and Variational Two-Electron Reduced-Density Matrix Methods}. \emph{Journal
  of Chemical Theory and Computation} \textbf{2019}, \emph{15}, 290--302\relax
\mciteBstWouldAddEndPuncttrue
\mciteSetBstMidEndSepPunct{\mcitedefaultmidpunct}
{\mcitedefaultendpunct}{\mcitedefaultseppunct}\relax
\EndOfBibitem
\bibitem[Meitei and Mayhall(2021)Meitei, and Mayhall]{Meitei2021}
Meitei,~O.~R.; Mayhall,~N.~J. {Spin-Flip Pair-Density Functional Theory: A
  Practical Approach to Treat Static and Dynamical Correlations in Large
  Molecules}. \emph{Journal of Chemical Theory and Computation} \textbf{2021},
  \emph{17}, 2906--2916\relax
\mciteBstWouldAddEndPuncttrue
\mciteSetBstMidEndSepPunct{\mcitedefaultmidpunct}
{\mcitedefaultendpunct}{\mcitedefaultseppunct}\relax
\EndOfBibitem
\bibitem[Dey and Ghosh(2022)Dey, and Ghosh]{Dey2022}
Dey,~M.; Ghosh,~D. {Curious Case of Singlet Triplet Gaps in Nonlinear
  Polyaromatic Hydrocarbons}. \emph{Journal of Physical Chemistry Letters}
  \textbf{2022}, \emph{13}, 11795--11800\relax
\mciteBstWouldAddEndPuncttrue
\mciteSetBstMidEndSepPunct{\mcitedefaultmidpunct}
{\mcitedefaultendpunct}{\mcitedefaultseppunct}\relax
\EndOfBibitem
\bibitem[Dhingra \latin{et~al.}(2023)Dhingra, Shori, and
  F{\"{o}}rster]{Dhingra2023}
Dhingra,~D.; Shori,~A.; F{\"{o}}rster,~A. {Chemically accurate singlet-triplet
  gaps of organic chromophores and linear acenes by the random phase
  approximation and {$\sigma$}-functionals}. \emph{Journal of Chemical Physics}
  \textbf{2023}, \emph{159}, 194105\relax
\mciteBstWouldAddEndPuncttrue
\mciteSetBstMidEndSepPunct{\mcitedefaultmidpunct}
{\mcitedefaultendpunct}{\mcitedefaultseppunct}\relax
\EndOfBibitem
\bibitem[Deleuze(2003)]{Deleuze2003TheTheories}
Deleuze,~M.~S. {The issues of size and charge consistency and the implications
  of translation symmetry in advanced Green's function theories}.
  \emph{International Journal of Quantum Chemistry} \textbf{2003}, \emph{93},
  191--211\relax
\mciteBstWouldAddEndPuncttrue
\mciteSetBstMidEndSepPunct{\mcitedefaultmidpunct}
{\mcitedefaultendpunct}{\mcitedefaultseppunct}\relax
\EndOfBibitem
\bibitem[Deleuze \latin{et~al.}(2003)Deleuze, Claes, Kryachko, and
  Fran{\c{c}}ois]{Deleuze2003BenchmarkOligoacenes}
Deleuze,~M.~S.; Claes,~L.; Kryachko,~E.~S.; Fran{\c{c}}ois,~J.~P. {Benchmark
  theoretical study of the ionization threshold of benzene and oligoacenes}.
  \emph{Journal of Chemical Physics} \textbf{2003}, \emph{119},
  3106--3119\relax
\mciteBstWouldAddEndPuncttrue
\mciteSetBstMidEndSepPunct{\mcitedefaultmidpunct}
{\mcitedefaultendpunct}{\mcitedefaultseppunct}\relax
\EndOfBibitem
\bibitem[Dupuy \latin{et~al.}(2015)Dupuy, Bouaouli, Mauri, Sorella, and
  Casula]{Dupuy2015VerticalAnsatz}
Dupuy,~N.; Bouaouli,~S.; Mauri,~F.; Sorella,~S.; Casula,~M. {Vertical and
  adiabatic excitations in anthracene from quantum Monte Carlo: Constrained
  energy minimization for structural and electronic excited-state properties in
  the JAGP ansatz}. \emph{Journal of Chemical Physics} \textbf{2015},
  \emph{142}, 214109\relax
\mciteBstWouldAddEndPuncttrue
\mciteSetBstMidEndSepPunct{\mcitedefaultmidpunct}
{\mcitedefaultendpunct}{\mcitedefaultseppunct}\relax
\EndOfBibitem
\bibitem[Rangel \latin{et~al.}(2016)Rangel, Hamed, Bruneval, and
  Neaton]{Rangel2016}
Rangel,~T.; Hamed,~S.~M.; Bruneval,~F.; Neaton,~J.~B. {Evaluating the GW
  Approximation with CCSD(T) for Charged Excitations Across the Oligoacenes}.
  \emph{Journal of Chemical Theory and Computation} \textbf{2016}, \emph{12},
  2834--2842\relax
\mciteBstWouldAddEndPuncttrue
\mciteSetBstMidEndSepPunct{\mcitedefaultmidpunct}
{\mcitedefaultendpunct}{\mcitedefaultseppunct}\relax
\EndOfBibitem
\bibitem[Jones \latin{et~al.}(1990)Jones, Guerra, Favaretto, Modelli, Fabrizio,
  and Distefano]{Jones1990Determination}
Jones,~D.; Guerra,~M.; Favaretto,~L.; Modelli,~A.; Fabrizio,~M.; Distefano,~G.
  Determination of the Electronic Structure of Thiophene Oligomers and
  Extrapolation to Polythiophene. \emph{Journal of Physical Chemistry}
  \textbf{1990}, \emph{94}, 5761--5766\relax
\mciteBstWouldAddEndPuncttrue
\mciteSetBstMidEndSepPunct{\mcitedefaultmidpunct}
{\mcitedefaultendpunct}{\mcitedefaultseppunct}\relax
\EndOfBibitem
\bibitem[Da~Silva~Filho \latin{et~al.}(2007)Da~Silva~Filho, Coropceanu, Fichou,
  Gruhn, Bill, Gierschner, Cornil, and
  Br{\'{e}}das]{DaSilvaFilho2007Hole-vibronicEnergies}
Da~Silva~Filho,~D.~A.; Coropceanu,~V.; Fichou,~D.; Gruhn,~N.~E.; Bill,~T.~G.;
  Gierschner,~J.; Cornil,~J.; Br{\'{e}}das,~J.~L. {Hole-vibronic coupling in
  oligothiophenes: Impact of backbone torsional flexibility on relaxation
  energies}. \emph{Philosophical Transactions of the Royal Society A:
  Mathematical, Physical and Engineering Sciences} \textbf{2007}, \emph{365},
  1435--1452\relax
\mciteBstWouldAddEndPuncttrue
\mciteSetBstMidEndSepPunct{\mcitedefaultmidpunct}
{\mcitedefaultendpunct}{\mcitedefaultseppunct}\relax
\EndOfBibitem
\bibitem[Leitner \latin{et~al.}(2024)Leitner, Dempwolff, and
  Dreuw]{Leitner2024Fourth-OrderMethods}
Leitner,~J.; Dempwolff,~A.~L.; Dreuw,~A. {Fourth-Order Algebraic Diagrammatic
  Construction for Electron Detachment and Attachment: The IP- and EA-ADC(4)
  Methods}. \emph{Journal of Physical Chemistry A} \textbf{2024}, \emph{128},
  7680--7690\relax
\mciteBstWouldAddEndPuncttrue
\mciteSetBstMidEndSepPunct{\mcitedefaultmidpunct}
{\mcitedefaultendpunct}{\mcitedefaultseppunct}\relax
\EndOfBibitem
\bibitem[Leitner \latin{et~al.}(2022)Leitner, Dempwolff, and
  Dreuw]{Leitner2022ThePropagator}
Leitner,~J.; Dempwolff,~A.~L.; Dreuw,~A. {The fourth-order algebraic
  diagrammatic construction scheme for the polarization propagator}.
  \emph{Journal of Chemical Physics} \textbf{2022}, \emph{157}, 184101\relax
\mciteBstWouldAddEndPuncttrue
\mciteSetBstMidEndSepPunct{\mcitedefaultmidpunct}
{\mcitedefaultendpunct}{\mcitedefaultseppunct}\relax
\EndOfBibitem
\bibitem[Loos and Jacquemin(2020)Loos, and Jacquemin]{Loos2020d}
Loos,~P.~F.; Jacquemin,~D. {Is ADC(3) as Accurate as CC3 for Valence and
  Rydberg Transition Energies?} \emph{Journal of Physical Chemistry Letters}
  \textbf{2020}, \emph{11}, 974--980\relax
\mciteBstWouldAddEndPuncttrue
\mciteSetBstMidEndSepPunct{\mcitedefaultmidpunct}
{\mcitedefaultendpunct}{\mcitedefaultseppunct}\relax
\EndOfBibitem
\bibitem[Alml{\"{o}}f(1991)]{Almlof1991}
Alml{\"{o}}f,~J. {Elimination of energy denominators in M{\o}ller-Plesset
  perturbation theory by a Laplace transform approach}. \emph{Chemical Physics
  Letters} \textbf{1991}, \emph{181}, 319--320\relax
\mciteBstWouldAddEndPuncttrue
\mciteSetBstMidEndSepPunct{\mcitedefaultmidpunct}
{\mcitedefaultendpunct}{\mcitedefaultseppunct}\relax
\EndOfBibitem
\bibitem[H{\"{a}}ser and Alml{\"{o}}f(1992)H{\"{a}}ser, and
  Alml{\"{o}}f]{Haser1992}
H{\"{a}}ser,~M.; Alml{\"{o}}f,~J. {Laplace transform techniques in
  M{\o}ller-Plesset perturbation theory}. \emph{The Journal of Chemical
  Physics} \textbf{1992}, \emph{96}, 489--494\relax
\mciteBstWouldAddEndPuncttrue
\mciteSetBstMidEndSepPunct{\mcitedefaultmidpunct}
{\mcitedefaultendpunct}{\mcitedefaultseppunct}\relax
\EndOfBibitem
\bibitem[Bintrim and Berkelbach(2021)Bintrim, and Berkelbach]{Bintrim2021}
Bintrim,~S.~J.; Berkelbach,~T.~C. {Full-frequency GW without frequency}.
  \emph{Journal of Chemical Physics} \textbf{2021}, \emph{154}, 041101\relax
\mciteBstWouldAddEndPuncttrue
\mciteSetBstMidEndSepPunct{\mcitedefaultmidpunct}
{\mcitedefaultendpunct}{\mcitedefaultseppunct}\relax
\EndOfBibitem
\bibitem[De~Dominicis and Martin(1964)De~Dominicis, and
  Martin]{DeDominicis1964}
De~Dominicis,~C.; Martin,~P.~C. {Stationary entropy principle and
  renormalization in normal and superfluid systems. II. Diagrammatic
  formulation}. \emph{Journal of Mathematical Physics} \textbf{1964}, \emph{5},
  31--59\relax
\mciteBstWouldAddEndPuncttrue
\mciteSetBstMidEndSepPunct{\mcitedefaultmidpunct}
{\mcitedefaultendpunct}{\mcitedefaultseppunct}\relax
\EndOfBibitem
\bibitem[Bickers and Scalapino(1989)Bickers, and Scalapino]{Bickers1989}
Bickers,~N.~E.; Scalapino,~D.~J. {Conserving approximations for strongly
  fluctuating electron systems. I. Formalism and calculational approach}.
  \emph{Annals of Physics} \textbf{1989}, \emph{193}, 206--251\relax
\mciteBstWouldAddEndPuncttrue
\mciteSetBstMidEndSepPunct{\mcitedefaultmidpunct}
{\mcitedefaultendpunct}{\mcitedefaultseppunct}\relax
\EndOfBibitem
\bibitem[Bickers(2004)]{Bickers2004}
Bickers,~N.~E. In \emph{Theoretical Methods for Strongly Correlated Electrons};
  S{\'{e}}n{\'{e}}chal,~D., Tremblay,~A.-M., Bourbonnais,~C., Eds.; Springer,
  2004; pp 237--296\relax
\mciteBstWouldAddEndPuncttrue
\mciteSetBstMidEndSepPunct{\mcitedefaultmidpunct}
{\mcitedefaultendpunct}{\mcitedefaultseppunct}\relax
\EndOfBibitem
\bibitem[Rohringer \latin{et~al.}(2012)Rohringer, Valli, and
  Toschi]{Rohringer2012}
Rohringer,~G.; Valli,~A.; Toschi,~A. {Local electronic correlation at the
  two-particle level}. \emph{Physical Review B} \textbf{2012}, \emph{86},
  125114\relax
\mciteBstWouldAddEndPuncttrue
\mciteSetBstMidEndSepPunct{\mcitedefaultmidpunct}
{\mcitedefaultendpunct}{\mcitedefaultseppunct}\relax
\EndOfBibitem
\bibitem[Ai \latin{et~al.}(2025)Ai, Zhai, T{\"o}lle, and Chan]{ai2025quantum}
Ai,~Y.; Zhai,~H.; T{\"o}lle,~J.; Chan,~G.~K. Quantum many-body linear algebra,
  Hamiltonian moments, and a coupled-cluster inspired framework. \emph{The
  Journal of Chemical Physics} \textbf{2025}, \emph{163}, 011101\relax
\mciteBstWouldAddEndPuncttrue
\mciteSetBstMidEndSepPunct{\mcitedefaultmidpunct}
{\mcitedefaultendpunct}{\mcitedefaultseppunct}\relax
\EndOfBibitem
\end{mcitethebibliography}
\begin{tocentry}
\includegraphics[width=\textwidth]{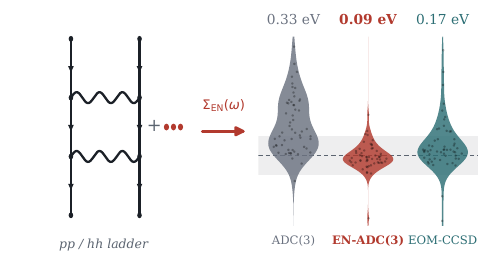}
\end{tocentry}

\end{document}